\documentclass[twocolumn]{aastex701}
\usepackage{natbib}
\usepackage[utf8]{inputenc}
\usepackage{booktabs}
\usepackage{color, amssymb, amsmath, natbib, color} %apjfonts,
\usepackage{multirow, placeins, subfigure}
\usepackage{graphicx}
\usepackage{epstopdf}
\usepackage{bm}
\usepackage{overpic}
\usepackage{hyperref}
\usepackage{comment}
\definecolor{lgray}{gray}{0.3}
\definecolor{orange}{rgb}{1,0.5,0}

\newdimen\digitwidth
\setbox1=\hbox{0}
\digitwidth=\wd1
\catcode`"=\active
\def"{\kern\digitwidth}

\newcommand\xone{x_1}
\newcommand\xtwo{x_2}
\newcommand\xthree{x_3}

\newcommand\C{{\mathcal C}}

\newcommand\simgt{\lower.5ex\hbox{$\; \buildrel > \over \sim \;$}}
\newcommand\simlt{\lower.5ex\hbox{$\; \buildrel < \over \sim \;$}}
\newcommand{\RNum}[1]{\uppercase\expandafter{\romannumeral #1\relax}}

\def\lzm{{\overline{L}_z}}
\def\rms{{R_\mathrm{rms}}}
\def\cam{\mathrm{CAM}}
\defcitealias{QS2021}{QS21}
\defcitealias{sormani_new}{H24}
\defcitealias{sormani2023}{S23}

\begin{document}

% Here are the title and abstract of this paper.

\title{Orbital classification in rotating bar potentials using an empirical proxy of the second integral of motion}

% authors and affiliation

\author[0000-0003-1309-3050]{Tian-Ye Xia}
\affiliation{Department of Astronomy, School of Physics and Astronomy, Shanghai Jiao Tong University\\ 
800 Dongchuan Road, Shanghai 200240, People's Republic of China;}
\email{xiaty_2020@sjtu.edu.cn}
%\affiliation{State Key Laboratory of Dark Matter Physics, School of Physics and Astronomy, Shanghai Jiao Tong University, Shanghai 200240, People's Republic of China}

%email: jtshen@sjtu.edu.cn}
%\affiliation{Key Laboratory for Particle Astrophysics and Cosmology (MOE) / Shanghai Key Laboratory for Particle Physics and Cosmology, Shanghai 200240, China}

\author[0000-0001-5604-1643]{Juntai Shen}
\affiliation{Department of Astronomy, School of Physics and Astronomy, Shanghai Jiao Tong University\\ 
800 Dongchuan Road, Shanghai 200240, People's Republic of China;}
\affiliation{State Key Laboratory of Dark Matter Physics, School of Physics and Astronomy, Shanghai Jiao Tong University, Shanghai 200240, People's Republic of China}
\affiliation{Key Laboratory for Particle Astrophysics and Cosmology (MOE) / Shanghai Key Laboratory for Particle Physics and Cosmology, Shanghai 200240, People’s Republic China}
\email{jtshen@sjtu.edu.cn}
\correspondingauthor{Juntai Shen}
\email[show]{jtshen@sjtu.edu.cn}

\author[0000-0001-6127-6957]{John Magorrian}
\affiliation{Rudolf Peierls Centre for Theoretical Physics, Beecroft Building, Parks Road, Oxford, OX1 3PU, UK}
\email{john.magorrian@physics.ox.ac.uk}
\correspondingauthor{John Magorrian}
\email[show]{john.magorrian@physics.ox.ac.uk}
%\correspondingauthor{John Magorrian}
%\email[show]{john.magorrian@physics.ox.ac.uk}

\author[0000-0003-3658-6026]{Yu-jing Qin}
\affiliation{Department of Astronomy and Astrophysics, Cahill Center for Astrophysics, California Institute of Technology, MC 249-17, 1200 E California Boulevard, Pasadena, CA 91125, USA}
\email{qinyj.astro@gmail.com}

\begin{abstract}

We present a novel method for classifying two-dimensional orbits in rotating bar potentials, based on an empirical proxy for the second integral of motion, Calibrated Angular Momentum ($\cam$), which is defined as the ratio of the time-averaged angular momentum ($\overline{L_z}$) to its temporal dispersion ($\sigma_{L_z}$) in the corotating frame. We show that $\cam$ is determined by the ratio of the azimuthal to radial actions (${J_\phi}^\prime / {J_r}^\prime$) in the analytical Freeman bar model. We then construct a new parameter space defined by $\cam$ versus the root-mean-square radius ($\rms$), and apply this framework to orbits in several representative rotating bar potentials. In the $\cam-\rms$ plane, periodic orbits generate well-defined branches separating distinct regions corresponding to different orbital families. Several of these branches enclose isolated areas that can be associated with specific orbital families, such as the the $x_2$ orbital family. We further validate the method using orbits from test-particle simulations, which show a well-ordered and non-overlapping distribution of orbital families in the $\cam-\rms$ plane. Since $\cam$ is fundamentally linked to intrinsic orbital properties and readily applied to three-dimensional orbits in $N$-body simulations, our results establish the $\cam-\rms$ plane as a robust and efficient framework for orbit classification in rotating bars that complements conventional methods.

\end{abstract}


\section{Introduction}

Stellar orbits constitute the fundamental building blocks of galactic dynamics. Numerous studies have explored the structure and evolution of orbital families in rotating bar potentials, as barred spiral galaxies account for nearly two-thirds of all spiral galaxies, including our Milky Way \citep{Binney1991,shen10,shenzheng2020}. 

In the two-dimensional (2D) case, the most important orbital families include the prograde $x_1$ orbits elonagated along the major bar axis, the retrograde $x_4$ orbits oriented perpendicular to the bar, as well as the prograde $x_2$ (stable) and $x_3$ (unstable) orbits, which are also perpendicular to the bar \citep{CG89, SW93, book1}. The $x_1$ family dominates the orbital population in self-gravitating bars, providing the primary support for the bar structure. In contrast, $x_4$ orbits are typically rounder in shape and are predominantly found at smaller radii \citep{Sparkesellwood1987, contopoulos}. $x_2$ and $x_3$ orbits exist only when an Inner Lindblad Resonance (ILR) exists. They could vanish in systems with low central density, rapid bar rotation, or particularly strong bars \citep{vs83, SS87, SW93,V2016}. 

In three-dimensional (3D) models, the overall picture of orbital families remains broadly consistent, especially for periodic orbits. The dominant contribution still arises from vertical bifurcations of the $x_1$ family, which extend the 2D orbital backbone into the vertical dimension. Several studies have demonstrated that 3D bars are composed primarily of these vertically extended $x_1$ descendants, supplemented by a few additional orbital families \citep{PF91, SW93, S2002a, S2002b}.

Conventional methods for orbital classification include the Poincaré surface of section (SoS), which projects orbital trajectories onto a 2D phase space plane. In this representation, different orbital families exhibit distinct patterns: periodic orbits appear as discrete points, regular orbits stay on invariant curves, chaotic orbits populate a diffuse `sea', and higher-order resonant orbits occupy small island-like structures between the regular and chaotic regions. Orbital families often form bull's eye patterns around parent orbits, enabling a qualitative classification of the orbital structure \citep{book2,SW93,shen04,book1}. 

However, despite its widespread recognition, the SoS method suffers from notable limitations. First, it is inherently energy-dependent and applicable only to orbits at a fixed energy, making it less suitable for analyzing orbits with a continuous energy distribution, as often encountered in $N$-body simulations. Second, it does not scale well to higher dimensions: in the 3D systems, extracting a meaningful 2D section from five-dimensional energy-constrained phase space becomes challenging and risks obscuring key dynamical features \citep{CA98}.

Another well-established diagnostic tool often used to aid orbital classification is frequency analysis, which can be applied to both 2D and 3D orbits \citep{BS82, laskar1993, CA98, VC07, V2010, V2016}. This technique constructs frequency maps by performing a spectral decomposition of orbits, where distinct orbital families appear as clusters of points centered around their resonant parent orbits. These resonances manifest as thin lines in the frequency space \citep{V2016}. Despite its utility, frequency analysis suffers from several limitations. First, the identification of fundamental frequencies can be technically challenging and computationally expensive. Second, frequency maps depend on the choice of coordinate system, which can cause resonant orbits to mix with others when different coordinates are used \citep{VC07, V2016}. Moreover, degeneracies can arise because different orbital families may share the same resonance. For example, the $x_2$ and $x_4$ orbits can both be associated with the $2:1$ radial-to-azimuthal resonance, making them difficult to be distinguished based solely on their spectral signatures.

To overcome these issues, we adopt an alternative classification framework based on an empirical proxy for the second integral of motion, originally introduced by \citet[hereafter \citetalias{QS2021}]{QS2021}. The second integral of motion ($I_2$) is a conserved quantity that constrains the phase space of regular orbits to a lower dimension. Unlike the Hamiltonian (i.e., the constant Jacobi energy, which combines the energy and angular momentum in the inertial frame through the relation $E_{\mathrm{J}}=E-\mathbf{\Omega_\mathrm{b}}\cdot \mathbf{L}$ in rotating bar potentials), $I_2$ typically lacks analytic expressions and is referred to as a non-classical integral of motion. Although these integrals can sometimes be inferred from the SoS, they remain difficult to isolate or define in general bar potentials \citep{book1}. To address this challenge, \citetalias{QS2021} proposed an empirical proxy for $I_2$ in rotating bar potentials, named as the Calibrated Angular Momentum (CAM). This quantity is defined as the ratio of the time-averaged angular momentum ($\overline{L_z}$) to its temporal dispersion ($\sigma_{L_z}$) in the corotating frame. Their analysis revealed that the often ignored physical quantity $\sigma_{L_z}$ actually contains valuable information, and that orbits with a constant Jacobi energy form a tight sequence in the $\overline{L_z}-\sigma_{L_z}$ plane. This sequence traces the main orbital families in the Poincaré SoS from $x_4$ to $x_1$ orbits, and accompanied by a monotonic variation in $\cam$, thereby demonstrating the effectiveness of $\cam$ as an empirical proxy for $I_2$.

The CAM framework offers several key advantages over traditional classification methods. First, it is effectively Hamiltonian-independent. By incorporating the Jacobi energy or other $E_\mathrm{J}$-related quantities as an additional parameter in the classification space, CAM can accommodate orbits spanning a broad range of Jacobi energies, making it inherently more suitable for analyzing realistic bar models than the SoS method, which is limited to iso-$E_\mathrm{J}$ slices. Moreover, in contrast to frequency-based techniques, CAM exhibits greater robustness against degeneracies among orbital families, as it does not rely on resonance structures that may overlap across different orbit types. Thirdly, the CAM framework is naturally extensible to 3D orbits. By incorporating an additional proxy for the third integral of motion, the method can potentially describe 3D orbits using just three parameters, which may provide a more visually intuitive alternative to SoS-based approaches.

Our main motivation is to investigate the effectiveness of $\cam$ for 2D orbital classification in rotating bar potentials. While \citetalias{QS2021} focused primarily on the empirical and numerical properties of $\cam$, we begin by investigating its analytical implications within the Freeman bar model, aiming to gain insight into the physical meaning of this proxy. Next, we evaluate the performance of $\cam$ for orbital classification across three representative rotating bar potentials: the Freeman bar, the logarithmic bar, and the Ferrers bar. In addition to the $\overline{L_z}-\sigma_{L_z}$ plane studied by \citetalias{QS2021}, we introduce a new parameter plane defined by $\cam$ and the root-mean-square radius ($\rms=\sqrt{\overline{R^2}}$)\footnote{Throughout this paper, we use $\rms$ to characterize the orbital size, but we do confirm that the mean absolute radius ($\overline{|R|}$) produces comparable results.}, {where the average is taken over time along the orbit using a uniform time interval}. We first explore the behavior of iso-$E_\mathrm{J}$ and periodic orbits within the $\cam$ framework, finding that periodic orbits form well-defined branches in the $\cam-\rms$ plane that effectively partition the space into domains corresponding to specific orbital families. We then extend the analysis to test-particle orbits rather than orbits constrained by fixed Jacobi energies, thereby demonstrating the applicability of this method to realistic $N$-body simulations. Our analyses show that different orbital families exhibit a well-ordered distribution in the $\cam-\rms$ plane. These results highlight CAM as a robust and efficient tool for orbital classification.

The paper is organized as follows. In Section~\ref{section: implication}, we review the definition of $\cam$ and examine its implications within the analytical Freeman bar model. Section~\ref{section: periodic orbit} introduces the $\cam-\rms$ plane and explores the behavior of iso-$E_\mathrm{J}$ and periodic orbits in this framework. In Section~\ref{section: method}, we extend the analysis to test-particle orbits. We discuss the broader implications of our results in Section~\ref{section: discussion}, and summarize in Section~\ref{section: conclusion}.
%CAM is verified to work well in generic rotating bar potentials, including potentials of self-consistent snapshot in a $N$-body bar simulation.

\section{Physical interpretation of CAM}
\label{section: implication}
 
CAM was first introduced by \citetalias{QS2021} as a numerical proxy of $I_2$ in a 2D rotating logarithmic bar potential. It is defined as the ratio of the time-averaged angular momentum in the bar corotating frame to its temporal dispersion, expressed as:
\begin{equation}
 \cam=\frac{\overline{L_z}}{\sigma_{L_z}},
\end{equation}

While $\cam$ has been numerically demonstrated to serve as an empirical proxy for $I_2$, its physical nature remains to be undertood. In this section, we investigate its physical meaning by expressing $\cam$ as a function of the orbital actions in a simple analytical bar model. We focus on the actions $\bm{J}$ because they are integrals of motion that form a complete set of canonical coordinates with angle variables $\bm{\theta}$ \citep{book1}. Moreover, some specific potentials admit analytical expressions for these actions, including the Freeman bar model that we adopt in this section.

The Freeman bar \citep{freeman1,freeman2} represents one of the simplest analytical bar models, in which all orbits can be expressed analytically. It produces a quadratic potential:
\begin{equation}
 \Phi(x,y)=\frac{1}{2}(\Omega_x^{2}x^{2}+\Omega_y^{2}y^{2}).
\end{equation}
Here we set $\Omega_x=1$ and $\Omega_y=\sqrt{2}$. The potential is rotating with a fixed pattern speed of $\Omega_{\rm{b}}=0.5$.

The effective potential of the Freeman bar is:

\begin{equation}
\begin{aligned}
\Phi_{\rm{eff}}&=\frac{1}{2}\Omega_x^{2}x^{2}+\frac{1}{2}\Omega_y^{2}y^{2}-\frac{1}{2}{\Omega_{\rm{b}}}^{2}(x^{2}+y^{2})\\
&=\frac{1}{2}\Phi_{xx}{x^{2}}+\frac{1}{2}\Phi_{yy}{y^{2}},
\end{aligned}
\end{equation}
where $\Phi_{xx}=\Omega_x^{2}-{\Omega_{\rm{b}}}^{2}$ and $\Phi_{yy}=\Omega_y^{2}-{\Omega_{\rm{b}}}^{2}$. Here, $(x, y)$ are Cartesian coordinates in the corotating frame.

Solving the equation of motion $\bm{\ddot{x}}=-\bm{\nabla}\Phi_{\rm{eff}}-2\bm{\Omega_{\rm{b}}}\times\bm{\dot{x}}$, the orbits of the Freeman bar model are \citep{freeman1}:

\begin{equation}
\begin{aligned}
&x=X_{\alpha}\cos\theta_\alpha+X_{\beta}\cos\theta_\beta,\\
&y=Y_{\alpha}\sin\theta_\alpha+Y_{\beta}\sin\theta_\beta,
\end{aligned}
\label{eq:orbit}
\end{equation}
where $\theta_\alpha=\alpha t +\phi_{\alpha}$ and $\theta_\beta=\beta t +\phi_{\beta}$, and where $\alpha$ and $\beta$ are the two positive roots ($\alpha^2<\Phi_{xx}<\Phi_{yy}<\beta^2$) of equation:
\begin{equation}
x^4-x^2(\Phi_{xx}+\Phi_{yy}+4{\Omega_\mathrm{b}}^2)+\Phi_{xx}\Phi_{yy}=0. 
\end{equation}

A general orbit in a Freeman bar is the superposition of two ellipses: the prograde $\alpha-$ellipse along the bar axis with frequency $\alpha$ (or $x_1$ orbits) and the retrograde $\beta-$ellipse perpendicular to the bar axis with frequency $\beta$ (or $x_4$ orbits). There is no $x_2$ or $x_3$ orbital families in the Freeman bar. 

The axial ratios $q_\alpha$ and $q_\beta$ of the $\alpha$-ellipse and the $\beta$-ellipse are two constants as below\footnote{$q_\alpha$ and $q_\beta$ correspond to $k_\alpha$ and $k_\beta$ in \citet{freeman1}, while $X_\alpha$ and $X_\beta$ correspond to $A_\alpha$ and $A_\beta$ in that paper.}:
\begin{equation}
\begin{split}
q_\alpha=\frac{Y_{\alpha}}{X_{\alpha}}=\frac{\Phi_{xx}-\alpha^2}{2\Omega_b\alpha}=\frac{2\Omega_b\alpha}{\Phi_{yy}-\alpha^2},\\
q_\beta=\frac{Y_{\beta}}{X_{\beta}}=\frac{\Phi_{xx}-\beta^2}{2\Omega_b\beta}=\frac{2\Omega_b\beta}{\Phi_{yy}-\beta^2},
\end{split}
\end{equation}
where $q_\alpha > 0$ and $q_\beta < 0$, indicating prograde motion for the $\alpha$-ellipse and retrograde motion for the $\beta$-ellipse, respectively. 

The angular momentum in the corotating frame is $L_z=x\dot{y}-y\dot{x}$, which can be rewritten as:
\begin{equation}
\begin{aligned}
%L_z=(X_1cos\theta_\alpha+X_2cos\theta_\beta)(\alpha{Y_1}cos\theta_\alpha+\beta{Y_2}cos\theta_\beta)+(Y_1sin\theta_\alpha+Y_2sin\theta_\beta)(\alpha{X_1}cos\theta_\alpha+\beta{X_2}cos\theta_\beta)
L_z=&\alpha{X_\alpha}{Y_\alpha}+\beta{X_\beta}{Y_\beta}+\\
&(\alpha q_\alpha+\beta q_\beta)X_\alpha X_\beta\cos\theta_\alpha{\cos\theta_\beta}+\\
&(\alpha q_\beta+\beta q_\alpha)X_\alpha X_\beta\sin\theta_\alpha{\sin\theta_\beta}.
\end{aligned}
\end{equation}
For nonzero $\Omega_{\mathrm{b}}$, the second and third terms cannot vanish simultaneously as $\alpha \ne \beta$, except in the case of periodic orbits for which $X_\alpha = 0$ or $X_\beta = 0$.

Thus, the time-averaged $L_z$ is:

\begin{equation}
\overline{L_z}=\alpha q_\alpha{X_\alpha}^2+\beta q_{\beta} X_\beta^2.
\end{equation}

The dispersion of $L_z$ can be obtained as:

\begin{equation}
\begin{aligned}
\sigma_{L_z}=&\left[
    \frac{1}{8}(\alpha+\beta)^2(q_\alpha+q_\beta)^2
\right. \\
& \left.
    + \frac{1}{8}(\alpha-\beta)^2(q_\alpha-q_\beta)^2
\right]^{\frac{1}{2}}
X_\alpha X_\beta.
\label{eq:sigmalz}
\end{aligned}
\end{equation}

Therefore, the relationship between $\overline{L_z}$ and $\sigma_{L_z}$ can be expressed as:
\begin{equation}
\frac{{\overline{L_z}}^2}{A^2}+\frac{{\sigma_{L_z}}^2}{B^2}=1,
\end{equation}
where $A$ and $B$ are constants that depend only on $\alpha$, $\beta$, $q_\alpha$ and $q_\beta$. 

Orbits under a fixed $E_\mathrm{J}$ satisfy:

\begin{equation}
E_{\rm{J}}=\frac{1}{2}{X_\alpha}^2(\alpha^2+\Phi_{yy}{q_\alpha}^2)+\frac{1}{2}{X_\beta}^2(\beta^2+\Phi_{yy}{q_\beta}^2).
\end{equation}

\begin{figure*}[!htbp]
  \centering
  \includegraphics[width=0.49\textwidth]{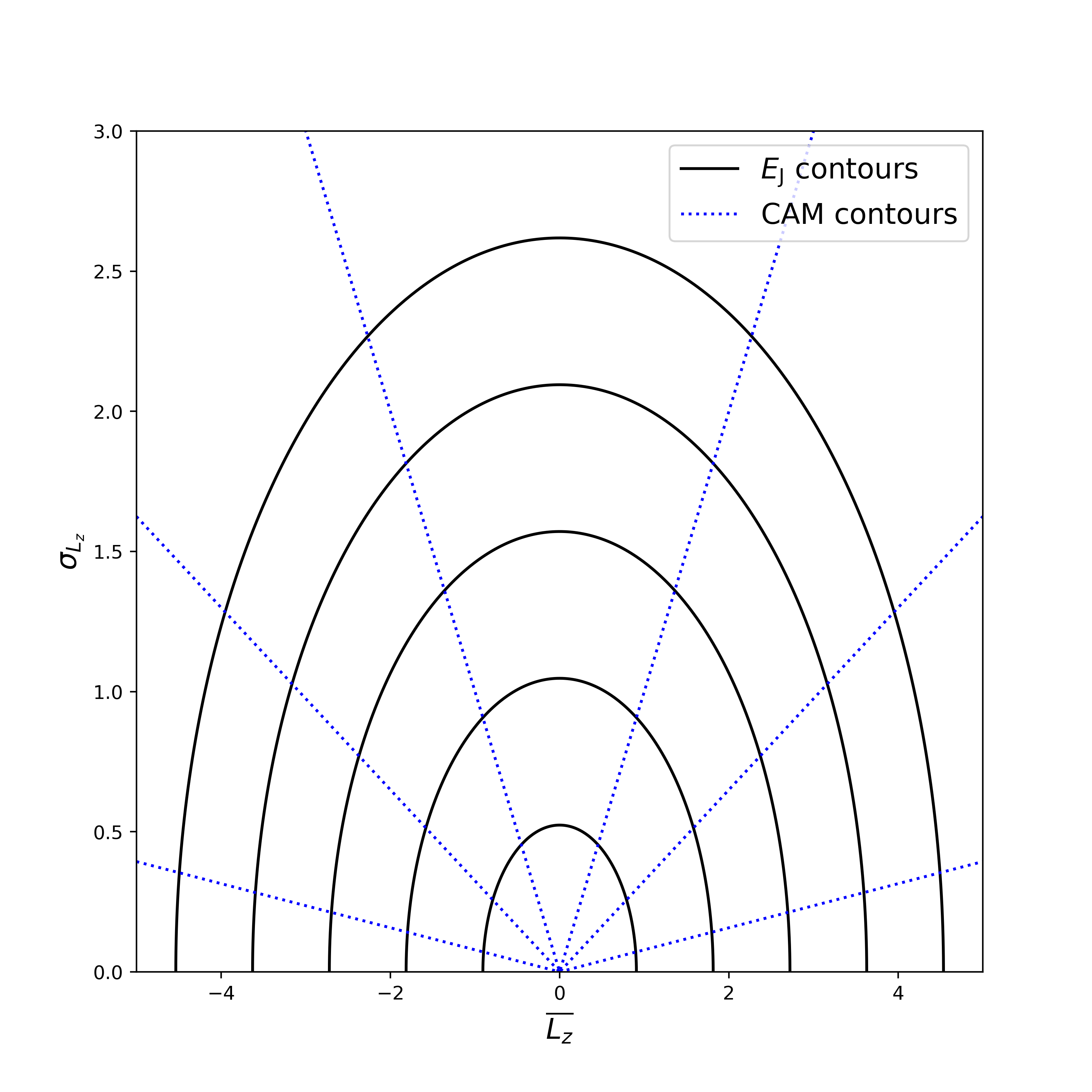},
  \includegraphics[width=0.49\textwidth]{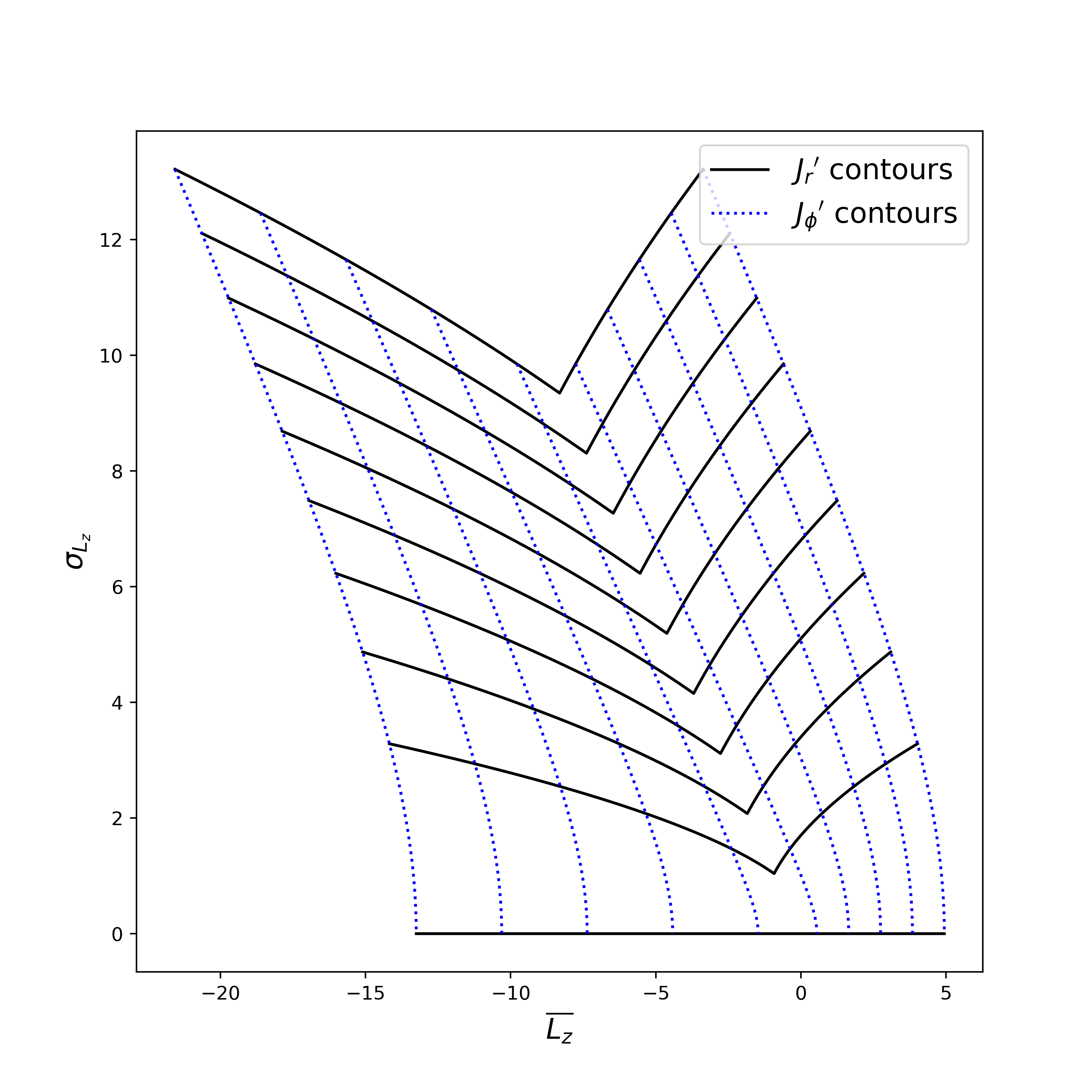}
  \caption{Contours of integrals of motion in the $\overline{L_z}-\sigma_{L_z}$ plane for orbits in the Freeman bar model. The left panel displays the $E_\mathrm{J}$ contours in black curves and the $\cam$ contours in blue dashed curves. The right panel shows the ${J_r}^\prime$ contours in black curves and the ${J_\phi}^\prime$ contours in blue dashed curves.}
  \label{fig:freeman_contour}
\end{figure*}

Thus, the $\overline{L_z}-\sigma_{L_z}$ distribution for orbits in a given Freeman bar potential at fixed $E_{\mathrm{J}}$ lies along the upper half of an ellipse, as shown by the black curves in the left panel of Figure~\ref{fig:freeman_contour}. Along this half-ellipse curve, from the bottom-left to the bottom-right, a clear transition from $x_4$ to $x_1$ orbits emerges, confirming that CAM traces the variations of $I_2$, as suggested in \citetalias{QS2021}. We also plot the $\cam$ contours in the blue dashed curves of the left panel of Figure~\ref{fig:freeman_contour}, which appear as straight rays emanating from the origin.

Since this potential corresponds to a 2D harmonic oscillator, the action-angle variables can be written simply as \citep{freeman1}:

\begin{equation}
\begin{split}
J_\alpha&=\frac{1}{2}(\alpha q_\alpha-\beta q_\beta)q_\alpha X_\alpha^2,\\
J_\beta&=-\frac{1}{2}(\alpha q_\alpha-\beta q_\beta)q_\beta X_\beta^2;\\
\theta_\alpha&=\alpha t +\phi_{\alpha},\\
\theta_\beta&=\beta t +\phi_{\beta}.\\
\end{split}
\label{eq:jajb}
\end{equation}

The actions $J_\alpha$ and $J_\beta$ are proportional to $X_\alpha^2$ and $X_\beta^2$, respectively, which in turn scale with the areas enclosed by the $\alpha-$ and $\beta-$ellipses in a given potential, thereby quantifying the extent of motion along each ellipse.

The Hamiltonian can be written as:
\begin{equation}
 H=\alpha J_\alpha+\beta J_\beta\equiv E_\mathrm{J}.
\end{equation}

$\cam$ can then be expressed as a function of $J_\alpha$ and $J_\beta$:
\begin{equation}
 \cam=\frac{\overline{L_z}}{\sigma_{L_z}}=C_1\frac{\alpha J_\alpha-\beta J_\beta}{\sqrt{J_\alpha J_\beta}},
\end{equation}
where $C_1$ is a constant.

While $J_\alpha$ and $J_\beta$ are actions defined specifically for the harmonic oscillator, we next seek an alternative set of actions associated with the radial and azimuthal motions, which can provide information applicable to more general potentials. We therefore introduce a new set of action–angle variables $({J_r}^\prime, {J_\phi}^\prime, {\theta_r}^\prime, {\theta_\phi}^\prime)$, defined through the generating function $S(\theta_\alpha,\theta_\beta,{J_r}^\prime, {J_\phi}^\prime)$:
\begin{equation}
\begin{split}
 S(\theta_\alpha,\theta_\beta,{J_r}^\prime, {J_\phi}^\prime)=&\theta_\alpha\left({J_r}^\prime+\frac{s+1}{2}{J_\phi}^\prime\right)+\\
 &\theta_\beta\left({J_r}^\prime+\frac{s-1}{2}{J_\phi}^\prime\right),
 \end{split}
\end{equation}
where $s=\mathrm{sgn}({J_\phi}^\prime)$. The new set of action–angle variables can then be written as:

\begin{equation}
\begin{split}
{J_r}^\prime&=\frac{1}{2}\left[J_\alpha+J_\beta-s(J_\alpha-J_\beta)\right],\\
{J_\phi}^\prime&=J_\alpha-J_\beta;\\
{\theta_r}^\prime&=\theta_\alpha+\theta_\beta,\\
{\theta_\phi}^\prime&=\frac{1}{2}\left[s(\theta_\alpha+\theta_\beta)+(\theta_\alpha-\theta_\beta)\right].\\
\end{split}
\label{eq:jrjphi}
\end{equation}

We adopt the notation ${J_r}^\prime$ and ${J_\phi}^\prime$ because these quantities reduce exactly to the radial and azimuthal actions, $J_r$ and $J_\phi$, in the axisymmetric limit ($\Omega_x = \Omega_y$). 

In the axisymmetric case, the radial action is defined by
\begin{equation}
J_r=\frac{1}{\pi}\int_{r_-}^{r_+}{v_r\mathrm{d}r},
\end{equation}
where $r_-$ and $r_+$ are the pericenter and apocenter radii. The azimuthal action $J_\phi$ corresponds to the angular momentum in the inertial frame, and is related to the angular momentum in the corotating frame through
\begin{equation}
J_\phi=L_z+\Omega_\mathrm{b}r^2, 
\end{equation}
where $r$ is the cylindrical radius. 

In the non-axisymmetric case, ${J_r}^\prime$ can be shown to be proportional to the area enclosed by the invariant curves in the $r-v_r$ SoS, obtained by slicing the four-dimensional phase space at $\phi$ fixed to a constant value\footnote{This relation holds only for closed invariant curves. In cases where invariant curves are split segments, the actions must be computed by other means \citep{Binney85, Xia_2021}.}, such that
\begin{equation}
{J_r}^\prime = \frac{1}{2\pi} \iint_{D} \mathrm{d}v_r\mathrm{d}r. 
\label{eq:sos}
\end{equation}
A similar calculation can be applied to ${J_\phi}^\prime$. We verified the equivalence of Equation~\ref{eq:jrjphi} and Equation~\ref{eq:sos} for three randomly-selected orbits in the Freeman bar, obtaining agreement at the level of $10^{-5}$.

%$({J_r}^\prime, {J_\phi}^\prime, {\theta_r}^\prime, {\theta_\phi}^\prime)$ can also be derived from $({J_r}, {J_\phi}, {\theta_r}, {\theta_\phi})$ in the axisymmetric case using perturbation theory \citep{book1}, as shown in Appendix~\ref{section: Freeman perturbatiom}. Note that the ${J_r}^\prime$ and ${J_\phi}^\prime$ obtained through this approach represent only an approximation to Equation~\ref{eq:jrjphi}, since higher-order terms are ignored during the process.

In this new set of action–angle variables, the Hamiltonian can be written as

\begin{equation}
 H=\frac{\alpha+\beta}{2}\left(2{J_r}^\prime+|{J_\phi}^\prime|\right)+\frac{\alpha-\beta}{2}{J_\phi}^\prime.
\end{equation}

We can then write $\cam$ as a function of ${J_r}^\prime$ and ${J_\phi}^\prime$:

\begin{equation}
\begin{split}
 \cam=\frac{\overline{L_z}}{\sigma_{L_z}}=&C_2\frac{{J_r}^\prime+\frac{\alpha-\beta}{\alpha+\beta}(2{J_r}^\prime+|{J_\phi}^\prime|)}{\sqrt{{J_r}^\prime({J_r}^\prime+|{J_\phi}^\prime|)}}\\
 =&C_2\frac{1+\frac{\alpha-\beta}{\alpha+\beta}(2+\mathrm{sgn}(l))}{\sqrt{1+|l|}},
\end{split}
\end{equation}
where $C_2$ is a constant, and $l\equiv {J_\phi}^\prime/{J_r}^\prime$. 

We plot the ${J_r}^\prime$ and ${J_\phi}^\prime$ contours in the $\overline{L_z}-\sigma_{L_z}$ plane, as shown in the right panel of Figure~\ref{fig:freeman_contour}. We find that $\overline{L_z}$ is primarily governed by the ${J_\phi}^\prime$ contours, which run approximately vertically with a slight slope, whereas $\sigma_{L_z}$ is mainly characterized by the ${J_r}^\prime$ contours, which lie roughly along the horizontal direction but twist around the center. Overall, $\cam$ is not strictly proportional to ${J_\phi}^\prime / {J_r}^\prime$, but rather a function of this ratio in the Freeman bar model.

To conclude, we find that in the Freeman bar model, $\overline{L_z}$ primarily reflects the contribution of ${J_\phi}^\prime$, which describes the azimuthal motion, whereas $\sigma_{L_z}$ is mainly determined by ${J_r}^\prime$, which characterizes the radial motion. Consequently, $\cam$ is determined by the ratio of ${J_\phi}^\prime / {J_r}^\prime$. In more general bar models, where the actions do not admit analytical expressions, it remains to be tested whether $\cam$ is strictly a function of this ratio. Nevertheless, the Freeman bar model provides valuable insight into the physical nature of $\cam$.

\section{Iso-$E_\mathrm{J}$ orbits and periodic orbits in the $\cam$ framework}
\label{section: periodic orbit}

To investigate how different orbital families are distributed within the $\cam$ framework, we begin by examining the behavior of iso-$E_\mathrm{J}$ orbits, which form regular and continuous sequences in the $\overline{L_z}-\sigma_{L_z}$ plane. We also present the distribution of periodic orbits, which serve as the parent orbits for their associated orbital families. These periodic orbits are fundamental in galactic dynamics, as quasi-periodic orbits (regular orbits that deviate from periodic ones) are generally trapped and oscillate around their corresponding parent periodic orbits.

In the Freeman bar potential, the iso-$E_\mathrm{J}$ orbits in the $\overline{L_z}-\sigma_{L_z}$ plane are shown in the black curves in the left panel of Figure~\ref{fig:freeman_contour}, forming the upper portion of an ellipse. The periodic $x_1$ and $x_4$ orbits in this model, however, exhibit zero angular momentum dispersion and constant $L_z$ as $X_\alpha$ or $X_\beta$ equals to $0$ (see Equation~\ref{eq:sigmalz}), causing their $\cam$ values to diverge. As a result, their properties cannot be meaningfully analyzed within the $\cam$ framework. For this reason, we also consider two additional representative rotating bar potentials: the logarithmic bar and the Ferrers bar in \citet{SW93}. The logarithmic bar model admits only the $x_1$ and $x_4$ orbital families, whereas the Ferrers bar potential also supports the $x_2$ orbital family. {Orbits in these two potentials are integrated for a timespan of around $50$ dynamical timescales using the python package {\tt Agama}, with a uniform time interval of $0.001$ dynamical timescales \citep{agama_2019}. We have verified that extending or shortening the integration time by a factor of a few does not affect the main results.} The periodic orbits are obtained by launching particles along the $x$-axis at fixed $E_\mathrm{J}$ and identifying those that return to their initial phase-space coordinates. 

In this analysis, we focus on low-order resonant periodic orbits located within the corotation radius, as these orbits predominantly contribute to the bar. Higher-order resonant orbits (with resonance order larger than 6) are excluded, as they are often associated with irregular or chaotic behavior and tend to produce complex and less interpretable structures in the parameter space.

In addition to the $\overline{L_z}-\sigma_{L_z}$ plane examined in the original CAM study \citetalias{QS2021}, we also explore alternative parameter spaces that directly incorporate the $\cam$ value. Since CAM serves as a proxy for $I_2$ in addition to $E_{\mathrm{J}}$, the $\cam-E_{\mathrm{J}}$ plane, resembling the integral-of-motion space, naturally provides a framework for orbital classification. However, because orbits inside and outside the corotation radius overlap in the $\cam-E_\mathrm{J}$ plane, an issue discussed in detail in Section~\ref{Section: log_pot} later, we instead adopt the $\cam-\rms$ plane, {where $\rms$ is defined as the time-averaged root-mean-square radius.} We find that this parameter space provides a clear separation between orbital families and serves as a critical diagnostic for orbital classification within the $\cam$ framework.

\subsection{$x_1$ and $x_4$ orbits}
\label{section:x1x4}

The logarithmic bar potential adopted in this study is identical to that used in \citetalias{QS2021}, which can be written as:
\begin{equation}
 \Phi(x,y)=\frac{1}{2}v_0^2\ln(R_c^2+x^2+\frac{y^2}{b^2}).
 \label{eq:log}
\end{equation}
We take $v_0=1$, $R_c=0.1$ and $b=0.84$, with a pattern speed of $\Omega_\mathrm{b}=1$. {The corotation radius and circular velocity at corotation are $r_{\rm CR} = 0.99$ and $V_{\rm CR} = 0.99$, respectively, owing to the nonzero value of $R_c$. Radii and velocities are therefore expressed in units of $1.01\ r_{\rm CR}$ and $1.01\ V_{\rm CR}$.} This model only supports $x_1$ and $x_4$ orbital families, while $x_2$ and $x_3$ orbital families are absent.

\begin{figure*}[htbp!]
  \centering
  \includegraphics[width=0.49\textwidth]{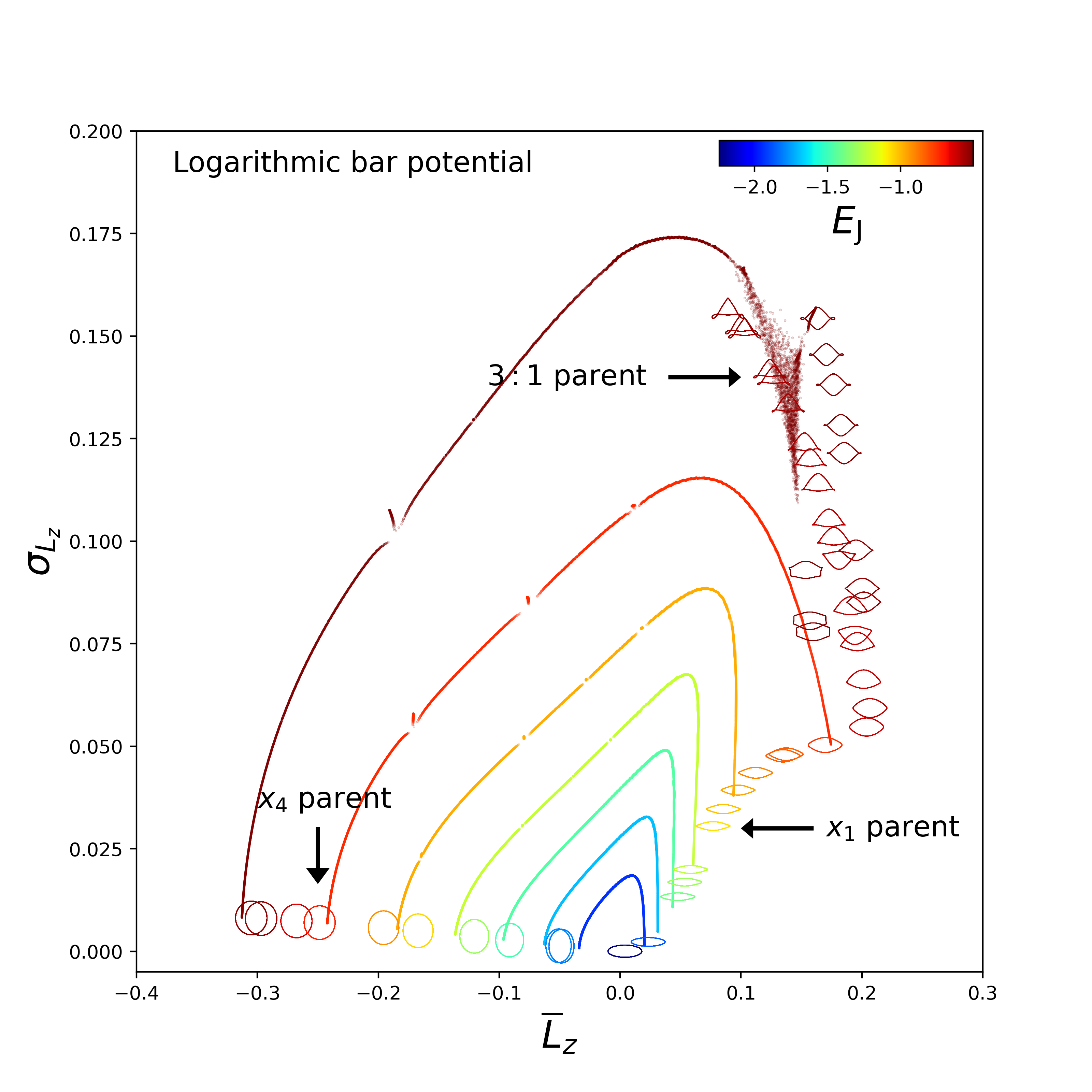}
  \includegraphics[width=0.49\textwidth]{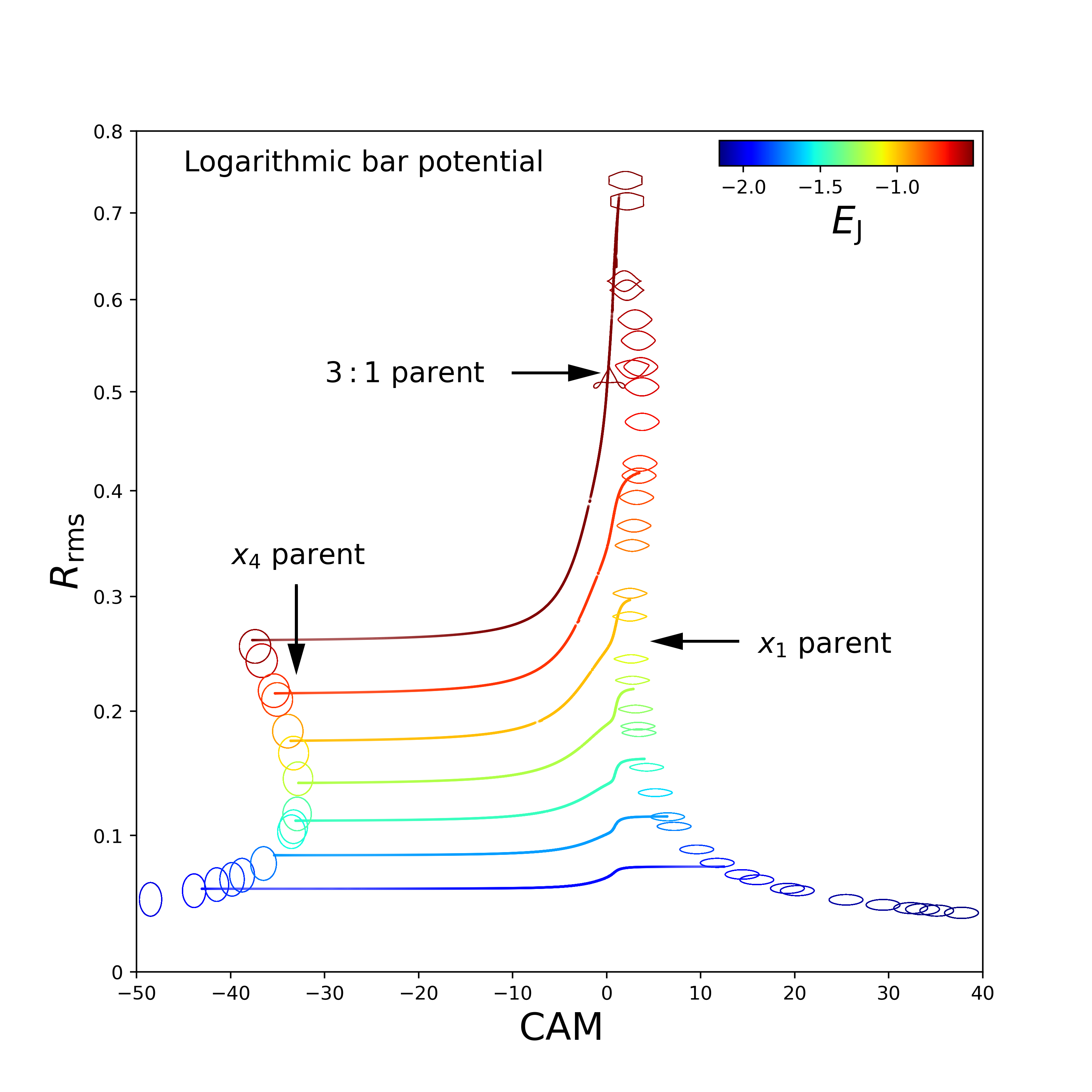}
  \caption{Orbital distributions of the logarithmic bar potential of Equation~\ref{eq:log}, shown in the $\lzm-\sigma_{L_z}$ plane (left) and the $\cam-\rms$ plane (right), color-coded by $E_{\mathrm{J}}$. Representative periodic orbits, displayed with their morphological shapes, are superimposed on the distribution, while quasi-periodic orbits appear as colored curves composed of discrete points. Distinct branches corresponding to periodic $x_1$, $x_4$ and 3:1 resonant orbits are labeled for reference. {The corotation radius is $1.01$, and the circular velocity and angular momentum at corotation are $1.01$ and $0.98$, respectively.}}
  \label{fig:log_lzsigmalz_peri}
\end{figure*}

Figure~\ref{fig:log_lzsigmalz_peri} shows the distribution of iso-$E_\mathrm{J}$ and periodic orbits of the logarithmic bar potential in the $\lzm-\sigma_{L_z}$ and $\cam-\rms$ planes, color-coded by $E_\mathrm{J}$. To facilitate comparison between periodic and quasi-periodic orbits, we display the scaled orbital shapes of representative periodic orbits, while quasi-periodic orbits at fixed Jacobi energies are indicated by curves made up of points. We plot the scaled periodic orbits using the following procedure: \romannumeral1) We divide the parameter space into a $20\times 20$ grid. \romannumeral2) We randomly select one periodic orbit from each populated grid cell. \romannumeral3) We plot the scaled shape of each sampled orbit at its corresponding coordinates in the parameter space. 

In the $\lzm-\sigma_{L_z}$ plane, shown in the left panel of Figure~\ref{fig:log_lzsigmalz_peri}, iso-$E_\mathrm{J}$ orbits form a continuous sequence from $x_4$ to $x_1$ orbits, as discovered by \citetalias{QS2021}. By comparing them with the periodic orbits, we find that low-$E_\mathrm{J}$ orbits are bounded by the $x_1$ and $x_4$ periodic orbits, which lie at the endpoints of the iso-$E_\mathrm{J}$ contours, as indicated by the labeled branches in the panel. These bounding periodic orbits approximately follow two distinct, nonlinear tracks, indicating that they correspond to different values of $\cam$. At higher $E_\mathrm{J}$, we observe a bifurcation of the $x_1$ periodic orbit into higher-order resonant families (e.g., the 3:1 resonance), giving rise to additional branches in the $\lzm-\sigma_{L_z}$ plane.

The right panel of Figure~\ref{fig:log_lzsigmalz_peri} shows the distribution of iso-$E_\mathrm{J}$ orbits and periodic orbits in the $\cam-\rms$ plane. We find that $\rms$ increases monotonically with $E_\mathrm{J}$ for orbits inside corotation radius, confirming the advantage of replacing $E_\mathrm{J}$ with $\rms$ as the $y$-axis of the parameter space.

We can clearly identify the morphology evolution for periodic orbits of different orbital families in the $\cam-\rms$ plane. For the $x_1$ orbits, low-$E_\mathrm{J}$ members initially appear as elongated ellipses. As $E_{\mathrm{J}}$ or $\rms$ rises, these orbits transition into lens-shaped configurations, followed by the development of less elongated and more elliptical shapes. At even higher $E_{\mathrm{J}}$, $x_1$ orbits display increasingly intricate morphologies, likely associated with higher-order resonant structures. For $x_4$ orbits, they evolve toward rounder shapes with growing $\rms$. 

Quasi-periodic orbits at fixed $E_\mathrm{J}$ also form well-defined curves in the $\cam-\rms$ plane, which represents the unfolded version of the iso-$E_\mathrm{J}$ contours in the $\lzm-\sigma_{L_z}$ plane. Along these curves, $\cam$ increases monotonically when progressing from $x_4$ to $x_1$ orbits, further supporting the validity of CAM as an effective proxy for $I_2$. This trend is accompanied by a mild variation in $\rms$, which exhibits a sharp increase near $\cam \sim 0$ but remains nearly constant otherwise. The variation of $\rms$ along each orbital sequence increases with $E_{\mathrm{J}}$, while the lowest-$E_{\mathrm{J}}$ orbits show minimal dispersion in $\rms$.

Similar to the $\lzm-\sigma_{L_z}$ plane, the $x_1$ and $x_4$ periodic orbits mark the boundaries for quasi-periodic orbits and trace two distinct curves. The two boundary lines do not maintain constant $|\cam|$ when $\rms$ increases: the absolute values of $\cam$ for the two boundaries shrink sharply at $\rms\lesssim0.1$, then vary more gradually at larger $\rms$.

\subsection{$x_2$ orbits}
\label{section: ferrers}
While the logarithmic bar potential discussed in Section~\ref{section:x1x4} supports only the $x_1$ and $x_4$ orbital families, in this section we adopt the Ferrers bar potential, which accommodates additional families such as $x_2$ orbits and more accurately represents the orbital structure of barred galaxies.

We adopt the bar potential described by \citet{SW93}, which comprises a \citet{ferrers1877} ellipsoid for the bar component, along with two spherically symmetric components, each following a Plummer density profile, representing the bulge and the halo.

The density profile of the bar component is given by:
\begin{equation}
\rho_{\rm B} = 
\begin{cases}
    \displaystyle \frac{105}{32\pi a c^2} M_{\rm B} (1 - \mu^2)^2, & \mu < 1, \\[6pt]
    0, & \mu \ge 1 ,
\end{cases}
\label{eq:rhoB}
\end{equation}
where
\begin{equation}
\mu^2 = \frac{x^2}{a^2} + \frac{y^2 + z^2}{c^2}.
\label{eq:mu}
\end{equation}
We take $M_B=1.1852$, $a=1$ and $c=\frac{1}{3}$, which are the same parameters used in \citet{SW93}.

The Plummer density profile for the spherically symmetric components is:
\begin{equation}
\rho_s=\frac{3M_s}{4\pi s^3}\left(1+\frac{r^2}{s^2}\right)^{-\frac{5}{2}}.
\label{eq:rhos}
\end{equation}
We adopt $M_s=0.3$ and $s=0.05$ for the spheroidal/bulge component, and $M_s=25$, $s=1.5$ for the halo component \citep{SW93}. The pattern speed is set to be $\Omega_\mathrm{b}=2$. {The corotation radius and the circular velocity at corotation are $r_{\rm CR} = 1.28$ and $V_{\rm CR} = 2.56$, respectively. All radii and velocities are expressed in units of $0.78\ r_{\rm CR}$ and $0.39\ V_{\rm CR}$.}

\begin{figure*}[htbp!]
  \centering
  \includegraphics[width=0.49\textwidth]{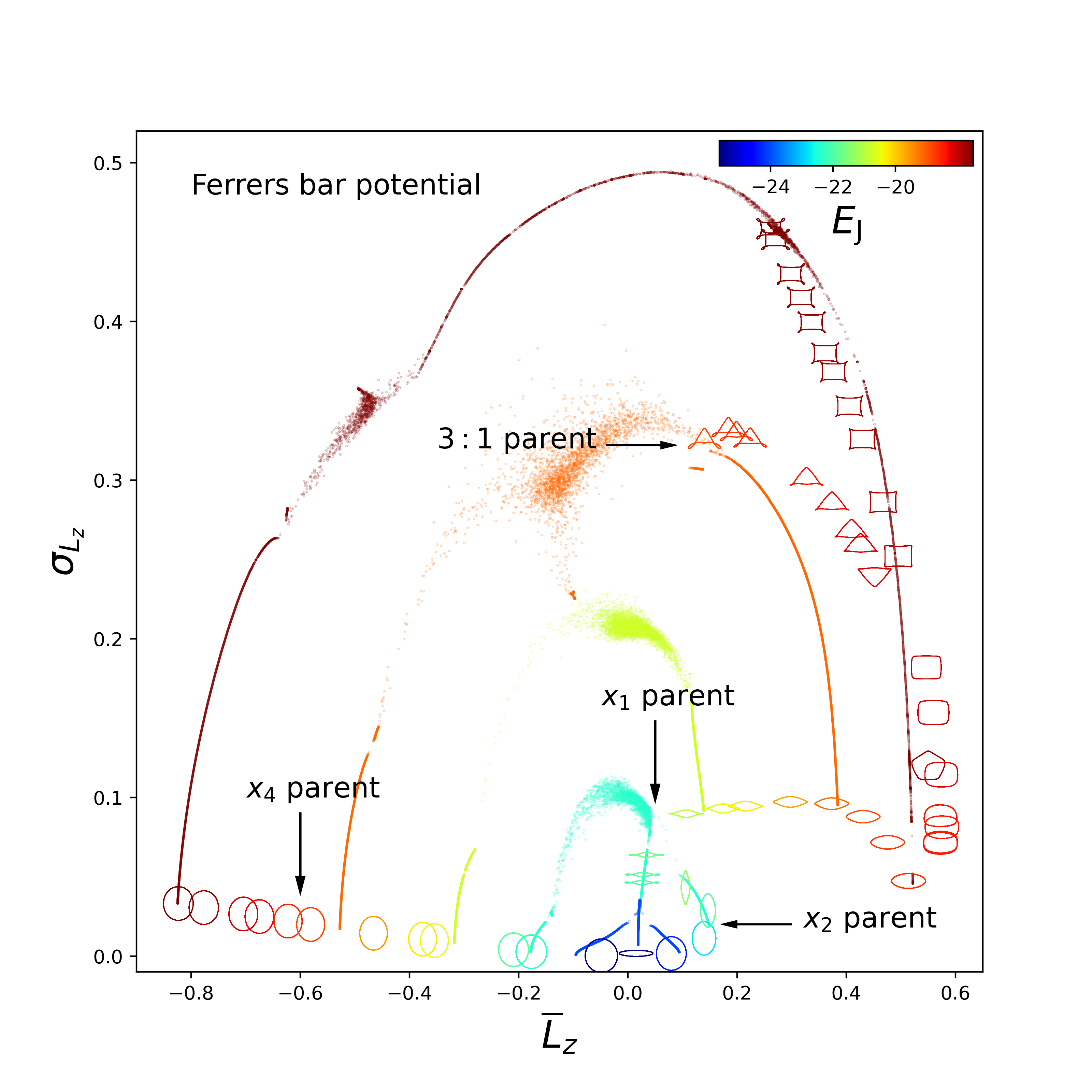}
  \includegraphics[width=0.49\textwidth]{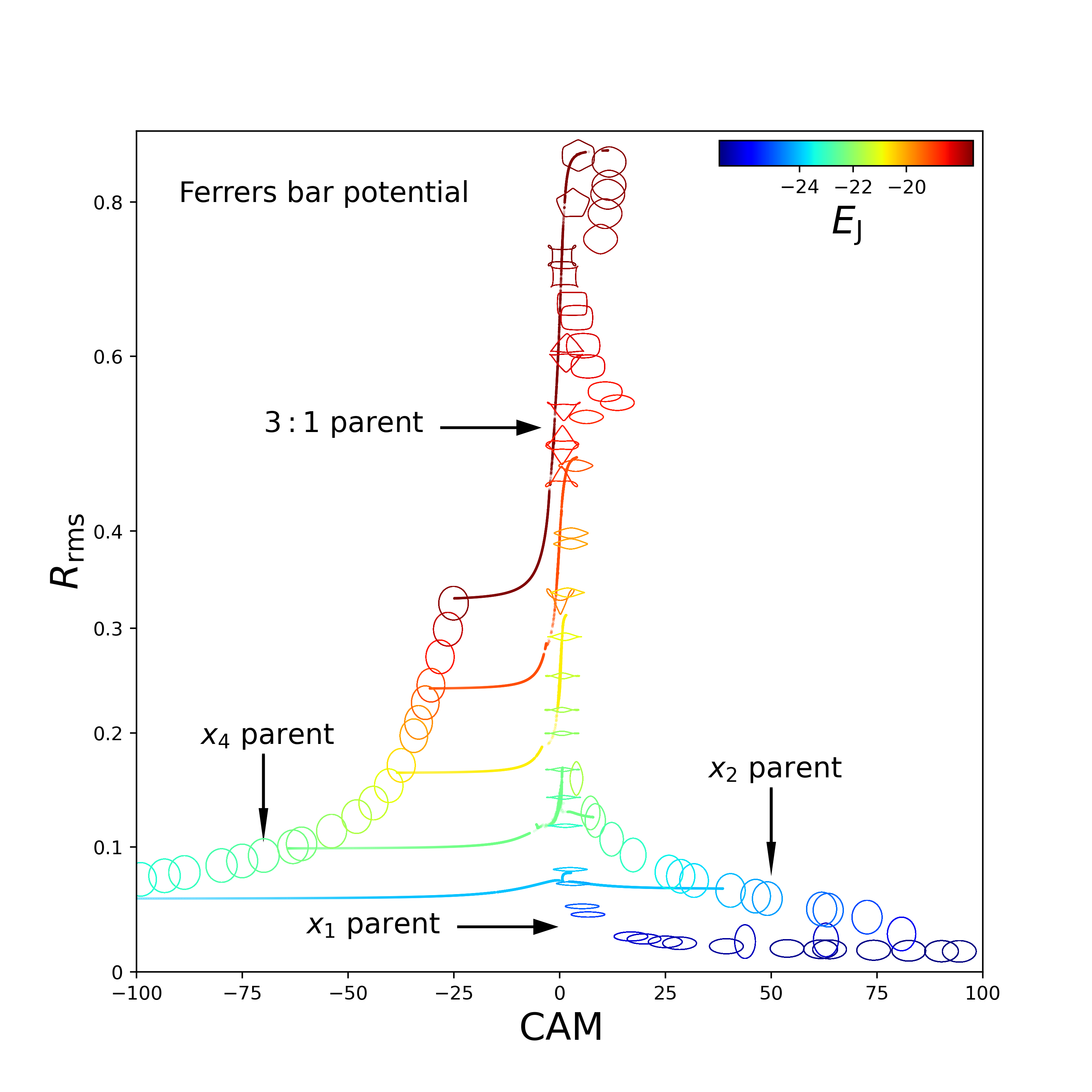}

  \caption{Same as Figure~\ref{fig:log_lzsigmalz_peri}, but for the Ferrers bar potential in Section~\ref{section: ferrers}. Chaotic are shown as scattered points, while quasi-periodic orbits appear as colored curves composed of discrete point. Distinct branches corresponding to periodic $x_1$, $x_2$, $x_4$ and 3:1 resonant orbits are labeled for reference. {The corotation radius is $1.28$, and the circular velocity and angular momentum at corotation are $2.56$ and $3.28$, respectively.}}
  %Orbital distributions of the Ferrers bar potential, shown in the $\lzm-\sigma_{L_z}$ plane (left) and the $\cam-\rms$ plane (right). Representative periodic orbits, displayed with their morphological shapes, are superimposed on the distribution, while quasi-periodic orbits appear as colored curves composed of discrete points. Chaotic orbits, in contrast, are shown as scattered points. Distinct branches corresponding to periodic $x_1$, $x_2$, $x_4$ and 3:1 resonant orbits are labeled for reference.}
  \label{fig:ferrer_lzsigmalz_peri}
\end{figure*}

Following the procedure outlined in Section~\ref{section:x1x4}, we map representative periodic orbits and iso-$E_\mathrm{J}$ orbits in the rotating Ferrers bar potential onto the $\lzm-\sigma_{L_z}$ and $\cam-\rms$ plane in Figure~\ref{fig:ferrer_lzsigmalz_peri}.

For the $\lzm-\sigma_{L_z}$ plane shown in the left panel of Figure~\ref{fig:ferrer_lzsigmalz_peri}, the periodic $x_1$ and $x_4$ orbits form two distinct branches that bound the quasi-periodic orbits similar to the case of the logarithmic bar potential. However, unlike the logarithmic potential where the quasi-periodic $x_1$-to-$x_4$ transition at a fixed $E_\mathrm{J}$ traces a smooth and continuous curve, orbits in the Ferrers bar potential exhibits a disruption in this trend. This disruption is caused by chaotic orbits scattered near $\overline{L}_z\sim0$, particularly at high $E_\mathrm{J}$.

In addition to the $x_1$ and $x_4$ orbital families, the $x_2$ orbits populate regions adjacent to the periodic $x_1$ branch at low Jacobi energies ($E_\mathrm{J}\lesssim-22$), characterized by slightly higher values of $\overline{L}_z$, as labeled in Figure~\ref{fig:ferrer_lzsigmalz_peri}. The periodic $x_2$ orbits form a distinct branch that is topologically connected to the periodic $x_1$ branch. The closed region bounded by the periodic $x_1$ and $x_2$ branches provides a distinct domain for $x_2$ orbits.

We show the distribution of iso-$E_\mathrm{J}$ and periodic orbits in the $\cam-\rms$ plane in the right panel of Figure~\ref{fig:ferrer_lzsigmalz_peri}. The contribution of chaotic orbits, which could disrupt the smoothness and continuity of iso-$E_\mathrm{J}$ sequences, is negligible, as chaotic orbits possess $|\cam|$ values that are significantly smaller than those of periodic orbits. Consequently, the overall continuity of the orbital sequence from $x_4$ to $x_1$ orbits is preserved, leading to smoother and more coherent iso-$E_\mathrm{J}$ sequences compared to those seen in the $\lzm-\sigma_{L_z}$ plane. This property enhances the reliability of orbital classification using the $\cam-\rms$ plane. At certain values of $\rms$, these trends extend further to reach the $x_2$ orbits. In these cases, however, the previously observed monotonic increase of $\rms$ with $\cam$ breaks down, and a decreasing trend in $\rms$ appears as the orbital family transitions from $x_1$ to $x_2$.

We find that the branch of periodic $x_2$ orbits is more prominently constructed in the $\cam-\rms$ plane than in the $\lzm-\sigma_{L_z}$ plane, while remaining topologically connected to the branch of periodic $x_1$ orbits. Together, the periodic $x_1$ and $x_2$ branches form a closed region that clearly delineates the $x_2$ orbital population. This structure provides a robust and visually intuitive framework for distinguishing between $x_1$ and $x_2$ orbits.

In addition, we find that some high-order resonance branches can also construct similar closed structures in the $\cam-\rms$ plane. For example, the periodic $3:1$ and $4:1$ resonance orbital branches, as well as the $x_1$ orbital branch at high $E_\mathrm{J}$, enclose well-defined regions. These closed regions may serve as useful indicators for identifying and classifying resonant orbital families.

To conclude, our analysis demonstrates that periodic orbits trace well-defined branches in the $\cam-\rms$ plane, effectively separating the parameter space into distinct regions occupied by quasi-periodic orbits of specific orbital families. This clear structure highlights the effectiveness of the $\cam-\rms$ plane as a robust framework for orbital classification.

\section{Test-particle orbits in the $\cam$ framework}
\label{section: method}

To more closely mimic real orbits in barred galaxies, we use orbits drawn from simple 2D test-particle simulations in three static bar potentials, including the Freeman bar, the logarithmic bar, and the Ferrers bar potentials. These orbits span a broader range of $E_\mathrm{J}$ and include the main orbital families associated with each bar model, thereby providing an overall view of the orbital distribution.

The simulation is set up by generating test particles with simple initial conditions as follows: \romannumeral1) The initial radius $R$ is drawn from an exponential distribution with scale length $r_d = 2.2$, such that the probability density is proportional to $\exp(-R/r_d)$. {We have verified that the specific choice of radial profile does not affect the final results.} The initial azimuthal angle $\phi$ is uniformly distributed between $0$ and $2\pi$. \romannumeral2) The initial velocities in the radial and azimuthal directions are sampled from different Gaussian distributions. The mean radial velocity is set to zero, while the mean azimuthal velocity is equal to the rotation velocity of the corresponding potential model. {The isotropic velocity dispersion is set to $0.4$ for the Freeman bar, $1$ for the Ferrers bar potentials, and $0.5$ for the logarithmic bar potential. These values are approximately half of the rotation velocity at corotation and are chosen to ensure a wide range of orbital behaviors, with the exception of the Freeman bar which does not possess a corotation radius. We further verify that varying the velocity dispersion by up to $50\%$ primarily affects the density of the orbital distribution, without altering its qualitative structure.} We integrate the orbits by using the python package {\tt Agama} \citep{agama_2019}. 

In addition to the $\cam-\rms$ plane, we also examine the distribution of test-particle orbits in the $\cam-E_{\mathrm{J}}$ plane, which resembles the integral-of-motion space. We verify that the $\cam-\rms$ plane provides a more effective tool for orbital classification.

\subsection{Freeman bar potential}
\label{Section: freeman}

The Freeman bar is indeed overly simplistic and idealized: it lacks a corotation radius and exhibits a rotation velocity that varies linearly with radius, which significantly deviates from the behavior of real barred galaxy potentials. Nevertheless, it can still serve as a useful local model for approximating bar orbits.

\begin{figure*}[!htbp]
  \centering
  \includegraphics[width=0.49\textwidth]{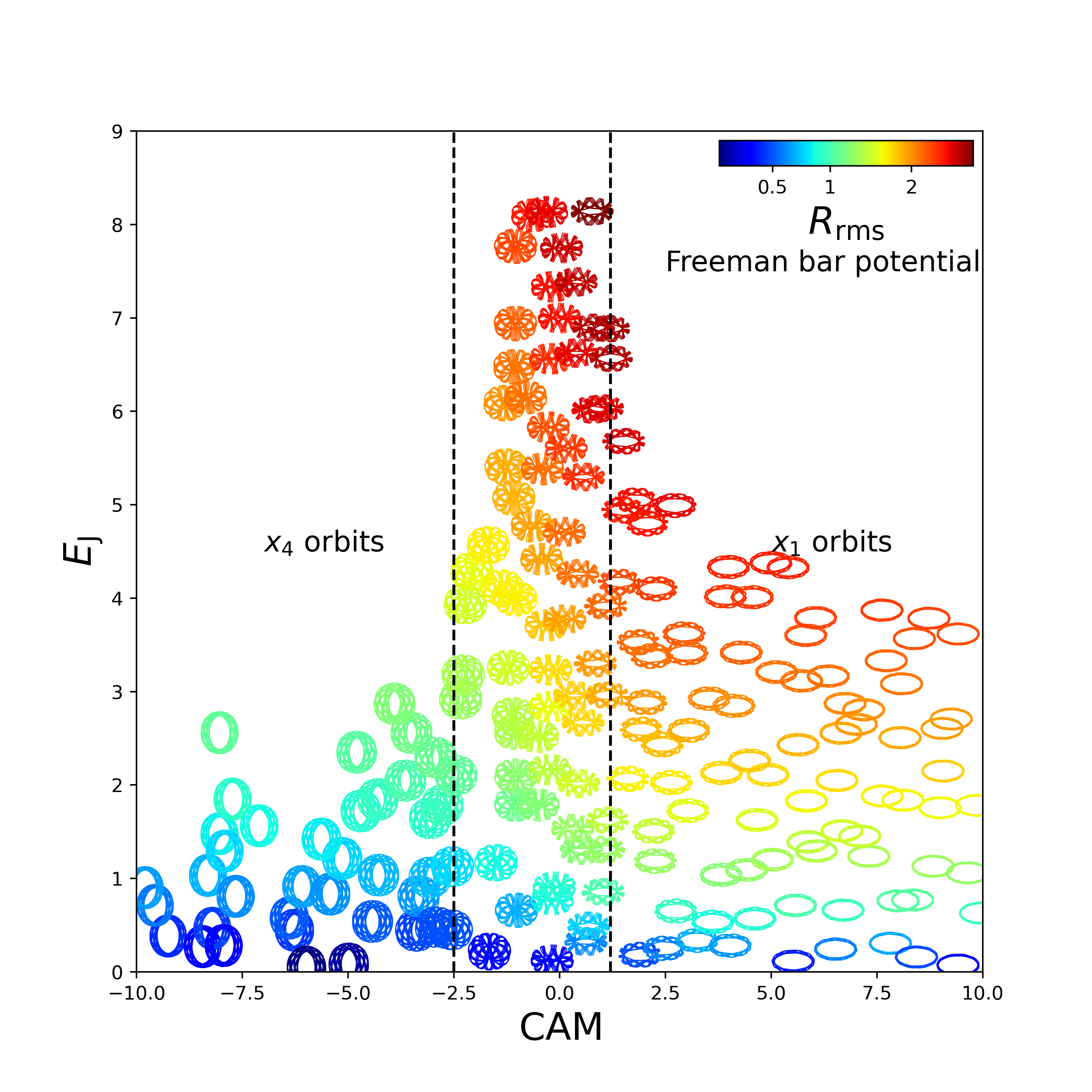}
  \includegraphics[width=0.49\textwidth]{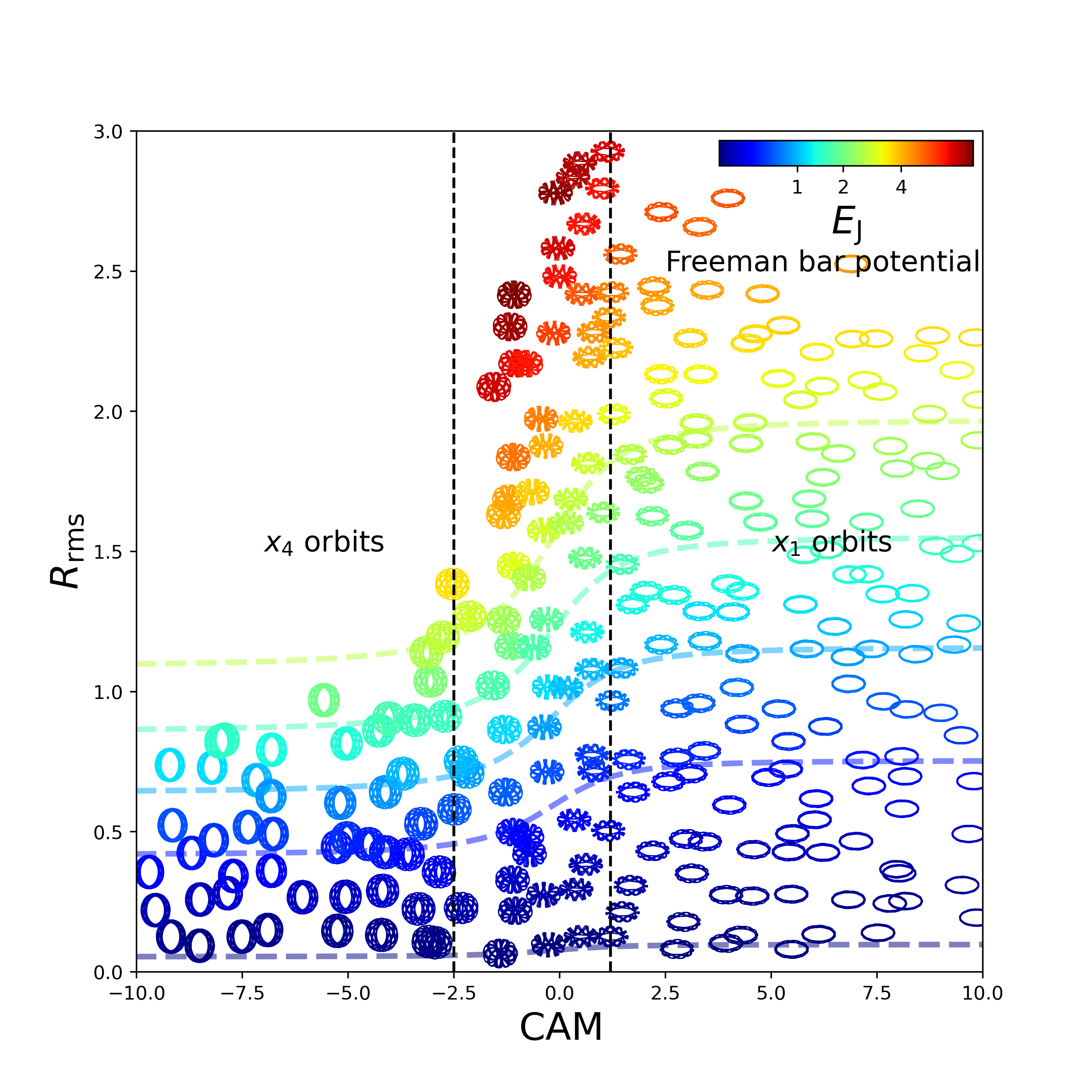}
  \caption{Orbital distribution in the Freeman bar model, shown in the $\cam-E_\mathrm{J}$ plane (left) and the $\cam-\rms$ plane (right). Scaled orbits are plotted at their corresponding locations in both parameter spaces. The left panel is color-coded by $\rms$, while the right panel is color-coded by $E_\mathrm{J}$. The approximate regions of the $x_1$, $x_2$ and $x_4$ orbital families are outlined by black dashed lines, and the colored dashed curves in the right panel trace the general trends of orbits at fixed $E_\mathrm{J}$. }
  \label{fig:freeman_cam_rms}
\end{figure*}

We plot the distribution of test-particle orbits for the Freeman bar potential in the $\cam-E_{\mathrm{J}}$ and $\cam-\rms$ planes, as shown in Figure~\ref{fig:freeman_cam_rms}. The panels are generated following the procedure described in Section~\ref{section:x1x4}.

We find a well-ordered transition from $x_4$ orbits to $x_1$ orbits in both the $\cam-E_{\mathrm{J}}$ and $\cam-\rms$ planes, where $x_4$ orbits occupy the left side of the diagrams, characterized by negative $\cam$, while $x_1$ orbits lie on the right with positive $\cam$. The parallel trends in these two planes arise from a smooth correlation between $E_{\mathrm{J}}$ and $\rms$. We label the approximate regions corresponding to the $x_1$ and $x_4$ orbital families, as there is no well-defined boundary between these two orbital families. In the left panel, the trend of orbits with similar spatial extents arises because $E_{\mathrm{J}}$ depends on angular momentum. In the right panel, we overplot iso-$E_\mathrm{J}$ sequences in this parameter space, shown as the colored dashed curves in the right panel of Figure~\ref{fig:freeman_cam_rms}, which are broadly consistent with the trends seen in the logarithmic bar potential (right panel of Figure~\ref{fig:log_lzsigmalz_peri}).

The noticeable shrinkage of the orbital-distribution boundaries as $E_\mathrm{J}$ or $\rms$ increases is caused by the initial conditions in our setup, which reduce the probability of generating orbits close to periodic orbits, which tend to exhibit large $|\cam|$. Although this behavior is common to our test-particle simulations for all rotating bar potentials, the Freeman bar potential is a special case: its periodic $x_1$ and $x_4$ orbits have formally infinite $|\cam|$, and therefore do not provide meaningful boundaries for the quasi-periodic orbits.

Overall, both the $\cam-E_{\mathrm{J}}$ and $\cam-\rms$ planes provide effective tools for orbit classification in the Freeman bar potential. Given the highly idealized nature of the Freeman bar, we therefore extend our analysis to more widely used models, the logarithmic and Ferrers bar potentials, to assess the applicability of these parameter spaces in more realistic settings.

\subsection{Logarithmic bar potential}
\label{Section: log_pot}

\begin{figure*}[htbp!]
  \centering
  \includegraphics[width=0.49\textwidth]{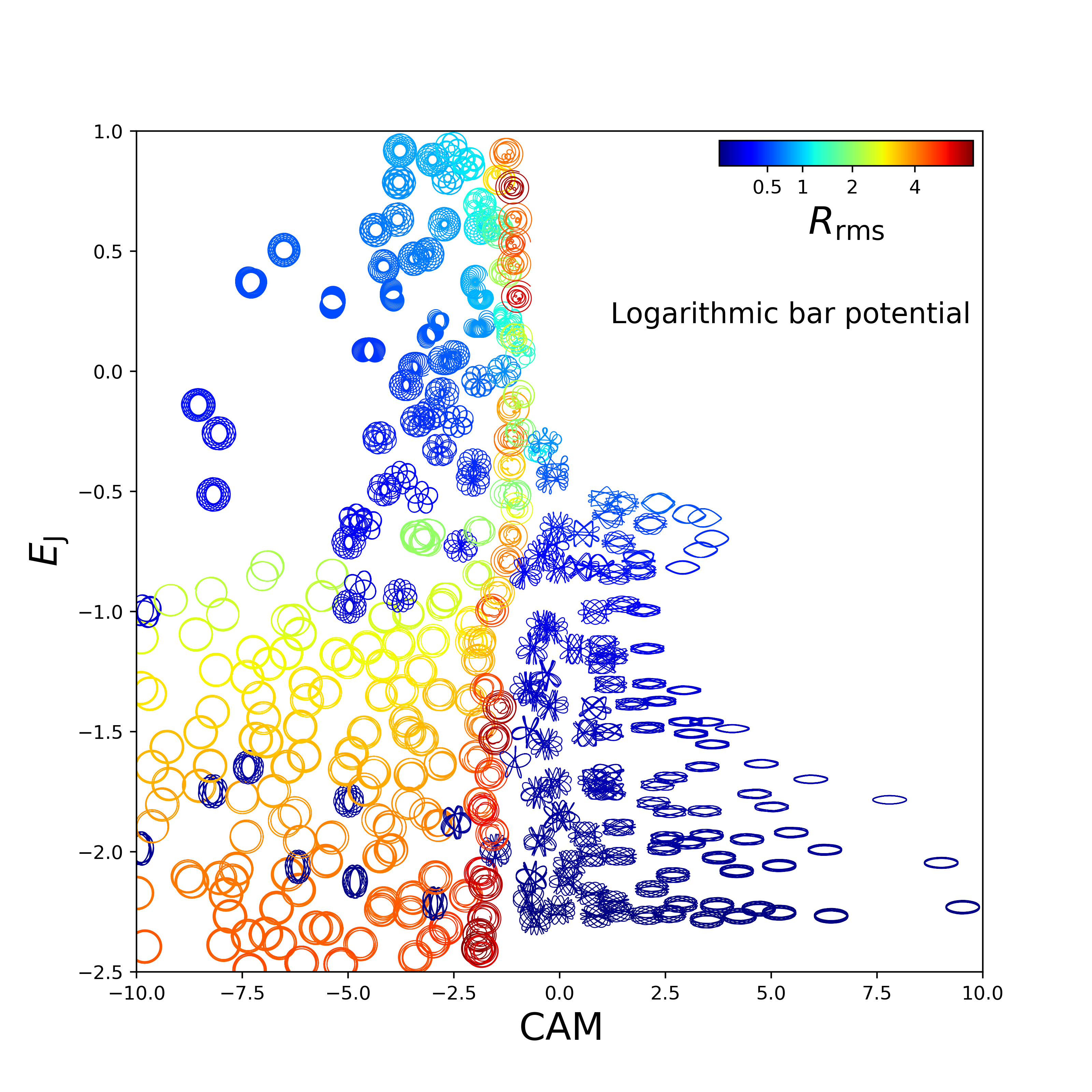}
  \includegraphics[width=0.49\textwidth]{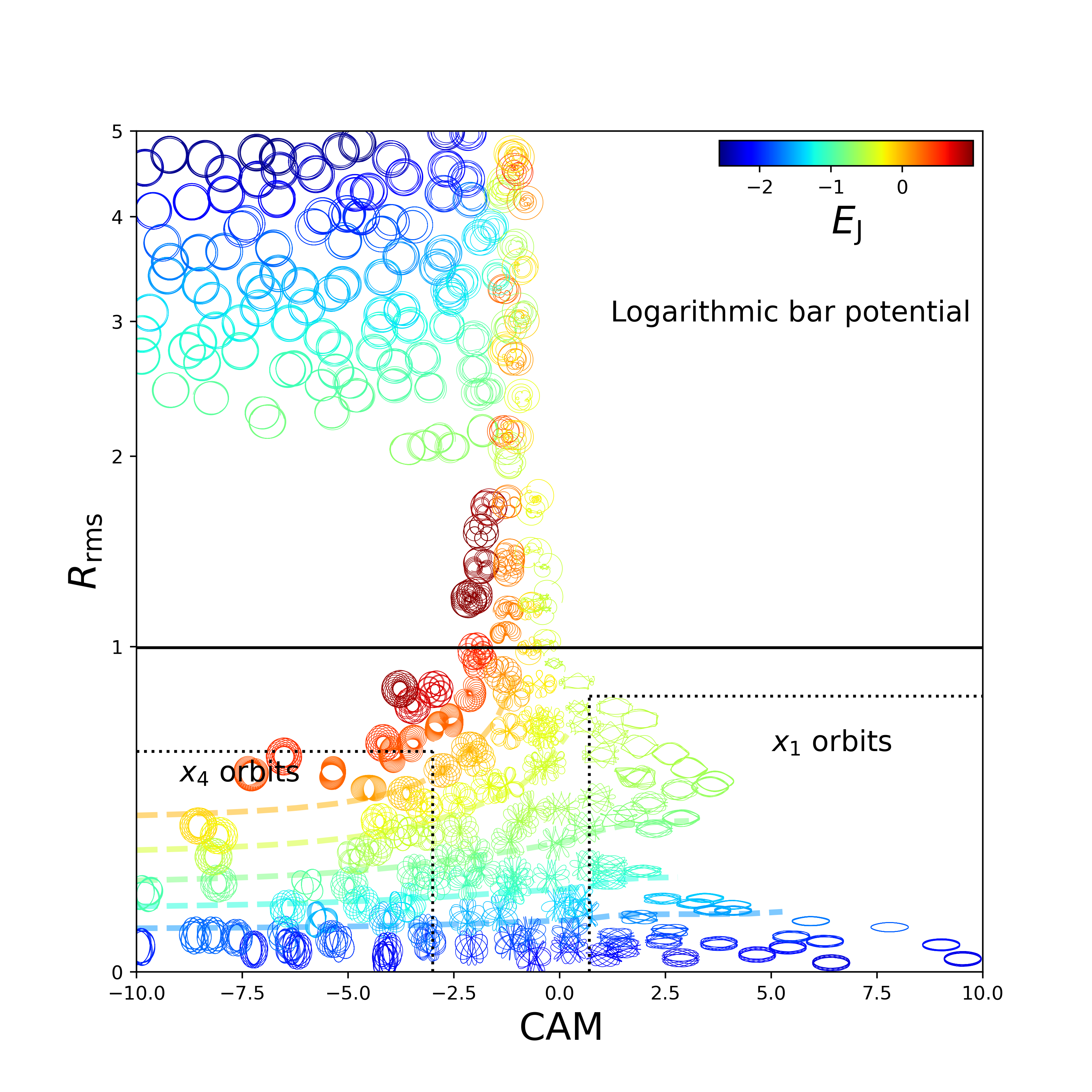}
  \caption{Same as Figure~\ref{fig:freeman_cam_rms}, but for the logarithmic bar potential. In the right panel, the approximate regions of the $x_1$, $x_4$ orbital families are outlined by black dashed lines, and the solid black horizontal line marks the corotation radius. The colored dashed curves trace the general trends of orbits at fixed $E_\mathrm{J}$. }
  \label{fig:log_cam_lz}
\end{figure*}

Similar to the case of the Freeman bar, we plot the orbital distribution of the logarithmic bar potential given by Equation~\ref{eq:log}, as well as the scaled shapes of representative orbits, in the $\cam-E_{\mathrm{J}}$ and $\cam-\rms$ planes in Figure~\ref{fig:log_cam_lz}. We find that the $x_1$ and $x_4$ orbital distributions in the this potential closely resemble those in the Freeman bar potential, where $x_1$ orbits occupy the lower-right region with $\cam >0$ and $x_4$ orbits are found in the lower-left region with $\cam <0$. This also agrees with the iso-$E_\mathrm{J}$ trend found by Figure~\ref{fig:log_lzsigmalz_peri}.

However, we notice that the region associated with the $x_4$ orbital family in the $\cam-E_{\mathrm{J}}$ plane is heavily populated by orbits located beyond the corotation radius. When color-coding orbits by $\rms$, we find that the continuous sequence from $x_4$ to $x_1$ orbital families becomes becomes disrupted by the presence of large-radius orbits. Orbits beyond the corotation radius, which rotate more slowly than the bar pattern speed and exhibit low $E_{\mathrm{J}}$ due to their large radius, introduce contamination into the $x_4$-to-$x_1$ sequence. These resulting degeneracies between orbits within and beyond corotation radius fundamentally limit the utility of $\cam-E_{\mathrm{J}}$ plane for reliable orbital classification. In contrast, such contamination does not occur in the Freeman bar model due to the absence of a defined corotation radius, as shown in the left panel of Figure~\ref{fig:freeman_cam_rms}. Nevertheless, the $\cam-E_{\mathrm{J}}$ plane remains a useful diagnostic for bar orbits once orbits outside the corotation radius are removed, as demonstrated in Appendix~\ref{appendix:insideco}.

In contrast, the $\cam-\rms$ plane effectively removes the degeneracies caused by the overlap of orbits inside and outside the corotation radius, yielding a clean distribution of orbital families. To highlight the progression of orbits at fixed $E_{\mathrm{J}}$, we fit cubic splines to the iso-$E_\mathrm{J}$ sequences in the right panel of Figure~\ref{fig:log_lzsigmalz_peri}, since iso-$E_\mathrm{J}$ orbits do not strictly form curves in logarithmic bar potentials, and plot the resulting trends as colored dashed curves in the right panel of Figure~\ref{fig:log_cam_lz}. In addition, we note that the $|\cam|$ values along the envelopes of the orbit population in the $\cam-\rms$ plane exhibit more pronounced variations than those in the right panel of Figure~\ref{fig:log_lzsigmalz_peri}, primarily because the two plots employ different ranges on the $\cam$ axis.

Within the corotation radius, the morphological evolution of quasi-periodic $x_1$ and $x_4$ orbits with increasing $\rms$ or $E_{\mathrm{J}}$ closely follows that of the parent periodic orbits in Figure~\ref{fig:log_lzsigmalz_peri}. When $\rms$ reaches the corotation radius, $x_4$ and $x_1$ orbits become trapped near the corotation region. In this regime, $x_4$ orbits transition into short-period orbits (SPOs), while $x_1$ orbits evolve into horse-shoe orbits, commonly referred to as long-period orbits (LPOs) \citep{SW93}.

Orbits beyond the corotation radius also exhibit well-ordered distributions in our analysis. These orbits consistently display negative $\cam$, as their rotation lags behind the bar pattern speed. Due to the averaging of the non-axisymmetric forces, they generally maintain nearly circular shapes and form narrow, rosette-like annuli with large $|\cam|$ values. {The outermost orbits, which have large $\rms$, correspond to the lowest values of $E_{\mathrm{J}}$ and are therefore more tightly bound. This behavior follows directly from the definition of the Jacobi energy: 
\begin{equation}
E_\mathrm{J}=\frac{1}{2}|\dot{\vec{r}}|^2+\Phi(\vec{r})-\frac{1}{2}\Omega_\mathrm{b}^2r^2,
\end{equation} 
where $\vec{r}$ and $\dot{\vec{r}}$ are defined in the corotation frame. For nearly circular orbits outside the corotation radius, the centrifugal term decreases more rapidly with increasing $\rms$ than the kinetic and gravitational potential terms, leading to lower $E_{\mathrm{J}}$ for larger $\rms$.}

In addition, we identify high-$E_\mathrm{J}$ orbits with $\cam\sim0$ in this region, primarily corresponding to unstable trajectories that would eventually escape the system under longer integration times.

\subsection{Ferrers bar potential}

\begin{figure*}[htbp!]
  \centering
  \includegraphics[width=0.49\textwidth]{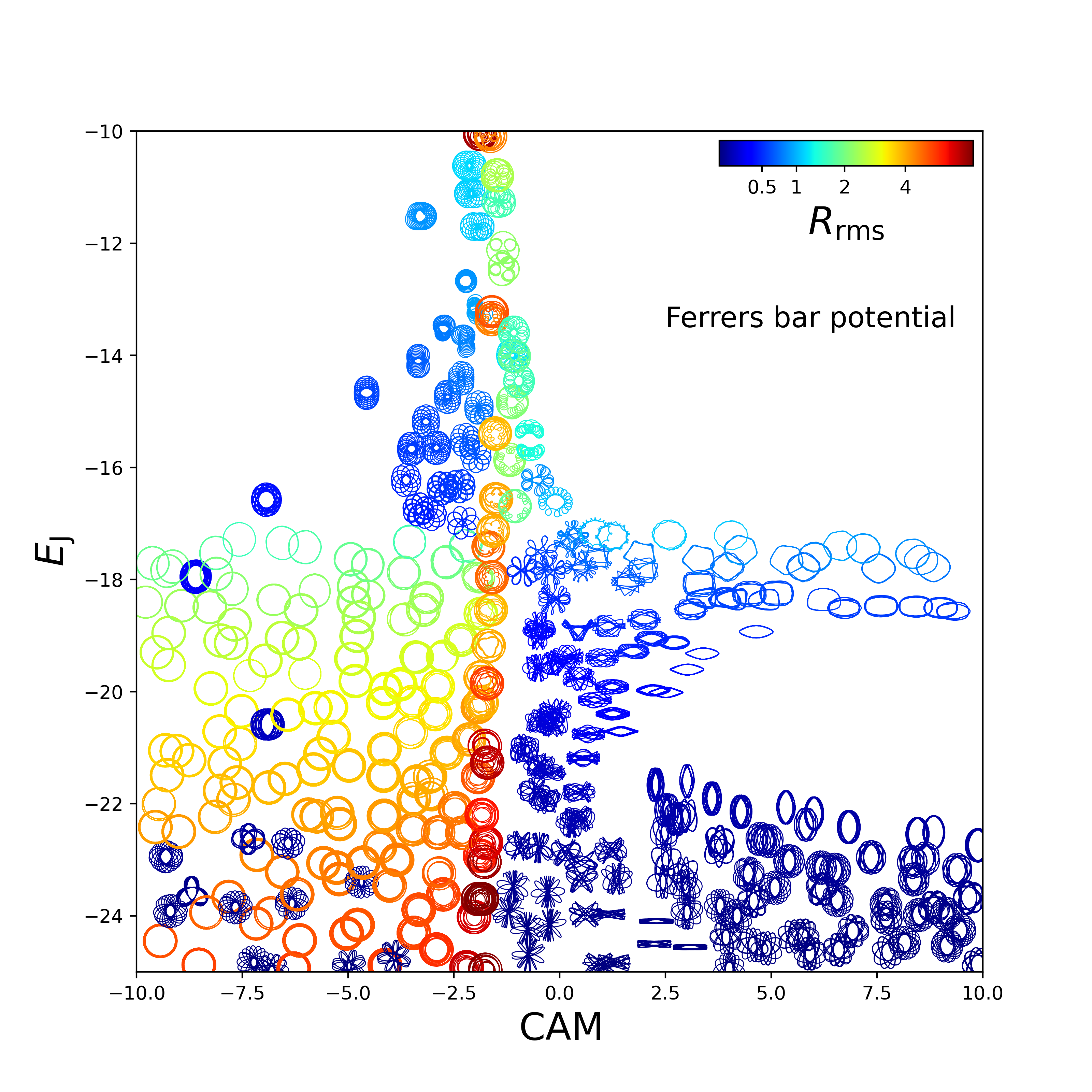}
  \includegraphics[width=0.49\textwidth]{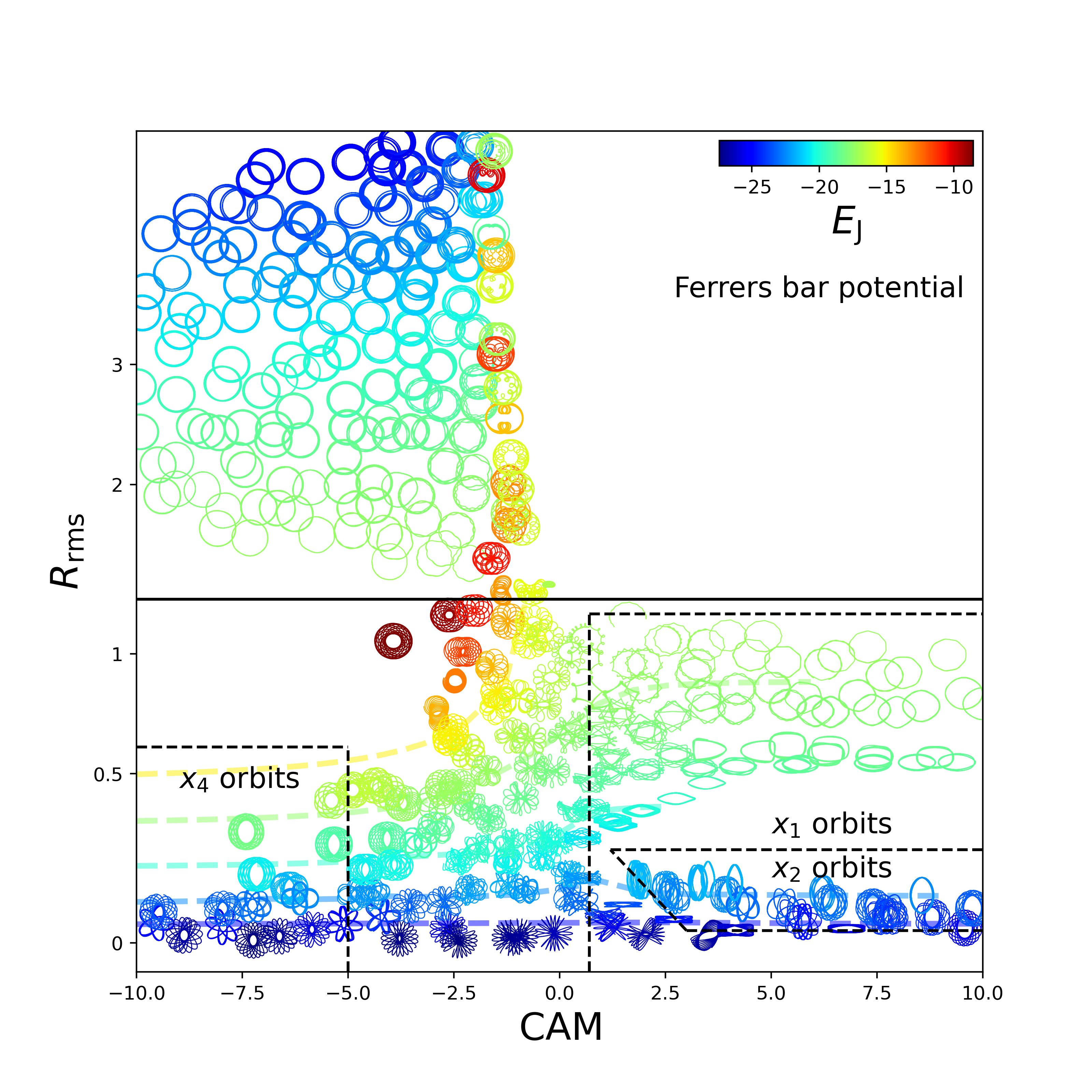}

  \caption{Same as Figure~\ref{fig:log_lzsigmalz_peri}, but for the Ferrers bar potential. The approximate regions of the $x_1$, $x_2$ and $x_4$ orbital families are outlined by black dashed lines. }
  \label{fig:ferrer_cam_lz}
\end{figure*}

To investigate the $x_2$ orbits, we map representative scaled orbits in the rotating Ferrers bar potential, of which the density profile is given by Equation~\ref{eq:rhoB} and Equation~\ref{eq:rhos}, onto the $\cam-E_{\mathrm{J}}$ and $\cam-\rms$ planes, as shown in Figure~\ref{fig:ferrer_cam_lz}.

Similar to the logarithmic bar potential, the $x_4$ orbits in the $\cam-E_{\mathrm{J}}$ plane are contaminated by orbits beyond the corotation radius, whereas a clean transition between the $x_4$ and $x_1$ orbital families within the corotation radius is revealed in the $\cam-\rms$ plane. As in the logarithmic bar case, we mark the fixed-$E_\mathrm{J}$ trends in Figure~\ref{fig:ferrer_cam_lz} with colored dashed lines, obtained by applying cubic-spline fits to the iso-$E_\mathrm{J}$ sequences in the right panel of Figure~\ref{fig:ferrer_lzsigmalz_peri}, after excluding chaotic orbits.

In particular, we find that the $x_2$ orbits emerge at low $\rms$ and high $\cam$, occupying a distinct region in the $\cam-\rms$ plane. These orbits are situated between the lens-shaped and boxy $x_1$ orbits at $0.05\lesssim \rms\lesssim 0.25$, and exhibit larger $\cam$ values than the $x_1$ orbits of comparable size, as depicted by the labeled regions for the $x_2$ orbital family in the right panel of Figure~\ref{fig:ferrer_cam_lz}. This location is consistent with the closed region bounded by the periodic $x_1$ and $x_2$ orbits in Figure~\ref{fig:ferrer_lzsigmalz_peri}. As a result, $x_2$ orbits show little overlap with the $x_1$ family in the $\cam-\rms$ plane, making this parameter space particularly effective for distinguishing between these two orbital families.

To conclude, the $\cam-\rms$ plane proves to be an effective and robust tool for orbital classification. It can be readily applied to snapshots of realistic $N$-body simulations, enabling the identification of major orbital families without requiring fixed-$E_\mathrm{J}$ sampling.

\section{Discussion}
\label{section: discussion}

\begin{comment}

\subsection{Orbits of high $E_\mathrm{J}$}
\label{section: high $E_\mathrm{J}$}

We notice that orbits under high $E_{\mathrm{J}}$ and inside corotation radius show complex behaviour in $\lzm-\sigma_{L_z}$ space. For example, as shown in the left panel of Figure~\ref{fig:ferrer_ej}, orbits of the Ferrers bar potential under a fixed $E_{\mathrm{J}}$ ($E_{\mathrm{J}}>-23$) become scattered around $\lzm=0$ in $\lzm-\sigma_{L_z}$ space, rather than confined a curve. This is because the existence of chaotic orbits among these $E_\mathrm{J}$ range.

However, we find that orbit distribution under high $E_\mathrm{J}$ becomes cleaner in the $\cam-\rms$ space, as shown in the right panel of Figure~\ref{fig:ferrer_ej}, which may due to the large range in the value of $\cam$. For those chaotic orbits that show scattering distribution in $\lzm-\sigma_{L_z}$ space around $\lzm=0$, they usually exhibit small value of $\cam$, and are thus less pronounced in the $\cam-\rms$ space.

\begin{figure*}[htbp!]
  \centering
  %\psfig{figs/lz_sigmalz.ps,width=0.5\textwidth}
  \includegraphics[width=0.44\textwidth]{ferrers_lzsigmalz_ej.png}
  \includegraphics[width=0.44\textwidth]{ferrers_camrms_ej.png}
  %\includegraphics[width=0.45\textwidth]{pl.pdf}
  \caption{$\lzm-\sigma_{L_z}$ and $\cam-\rms$ space of orbits under several fixed $E_{\mathrm{J}}$. }
  \label{fig:ferrer_ej}
\end{figure*}
\end{comment}

\subsection{Strengths and weaknesses}
\label{section: comparison}

Compared with conventional orbital classification tools, the $\cam$ diagnostic exhibits several key advantages. First, as a numerical proxy for the second integral of motion, $\cam$ is intrinsically independent of the Jacobi energy. It retains much of the dynamical information typically extracted from SoS, but without requiring orbits to be analyzed at fixed $E_{\mathrm{J}}$. This makes $\cam$ particularly well suited for studying test particles in barred galaxy models, where orbits span a continuous range of Jacobi energies, in contrast to the SoS, which is restricted to discrete $E_\mathrm{J}$-slices. Second, $\cam$ constrains orbital structure through phase-space averaging rather than direct projection, enabling a natural extension to 3D orbits through the incorporation of an additional proxy for the third integral of motion. Finally, because $\cam$ directly reflects intrinsic orbital properties as an integral-like quantity, it exhibits reduced degeneracy compared with frequency analysis, with distinct orbital families occupying non-overlapping regions.

Despite these advantages, we list its potential limitations below. 

The primary limitation of our method lies in the lack of quantitative criteria for distinguishing orbital families across different bar potentials. Because the locations of periodic-orbit branches can vary systematically among models, the mapping between orbital families and their corresponding regions in the $\cam-\rms$ plane is not universally defined. This issue becomes more pronounced for higher-order resonant orbits, whose domains in the $\cam-\rms$ plane may connect with multiple families. As a result, additional computational effort is required to solve for the relevant periodic orbits in each potential.

Another limitation arises from the reduced effectiveness of the method in identifying chaotic orbits, particularly at higher $E_\mathrm{J}$ where such orbits become more prevalent. This shortcoming stems from the fact that chaotic orbits do not preserve a well-defined second or third integral of motion, which undermines the applicability of CAM as a numerical proxy. Although chaotic orbits can be approximately located in the $\overline{L_z}$–$\sigma_{L_z}$ plane, their boundaries, unlike those of quasi-periodic orbits, cannot be anchored to periodic-orbit curves and therefore remain difficult to define rigorously. Further work is needed to develop quantitative criteria for identifying and isolating chaotic components within this framework.

\subsection{Application to 3D orbits}

Extending this framework to 3D orbital classification requires identifying an additional proxy that effectively captures the third integral of motion, $I_3$. Following the approximation schemes employed in the 2D case, a promising candidate we identified is the ratio of the time-averaged vertical action ($\overline{J_z}$) to the temporal dispersion of the angular momentum ($\sigma_{L_z}$). The vertical action is defined as $J_z=\int_{z_\mathrm{min}}^{z_\mathrm{max}}v_z\mathrm{d}z$, where $z_\mathrm{max}$ and $z_\mathrm{min}$ denote the turning points of a single vertical oscillation. We find that 3D orbits at fixed $E_\mathrm{J}$ trace a similar sequence in the $\overline{L_z}-\sigma_{L_z}$ plane to the 2D case, when they are grouped into bins of $\overline{J_z}/\sigma_{L_z}$, suggesting that this ratio may serve as a possible proxy for $I_3$. Therefore, the $\cam-\rms$ plane for 3D orbits can be constructed within each $\overline{J_z}/\sigma_{L_z}$ bin, potentially enabling effective 3D orbital classification. A detailed investigation and validation of these approaches will be presented in a forthcoming study.

%We have identified several promising candidates for this purpose, and a detailed investigation and validation of these approaches will be presented in a forthcoming study.
\subsection{Application to $N$-body simulations}
\label{section:n-body}

Here we briefly clarify how the method can be applied to particles in a realistic $N$-body simulation snapshot. For a given snapshot, one can extract the phase-space coordinates $(\mathbf{x}_n, \mathbf{v}_n)$ for particles, together with the frozen gravitational potential associated with the snapshot. To classify the orbital families represented by these particles, the following steps can be adopted:

i) {Integrate snapshot particles in the corresponding frozen potential. The quantities $\overline{L_z}$, $\sigma_{L_z}$, and $\rms$ can be computed by time-averaging along the resulting orbits, and $\cam$ is defined as the ratio $\overline{L_z}/\sigma_{L_z}$.}

ii) Construct the $\cam-\rms$ plane. The corotation radius that separates bar-supporting orbits from those outside corotation radius can be identified from the location in $\rms$ where a sharp decrease in the number of orbits at positive $\cam$ occurs (as illustrated by the black solid line in the right panel of Figure~\ref{fig:log_cam_lz}). Approximate clusters associated with different orbital families may then be identified by overplotting representative scaled orbits on this diagram, as shown in Figures~\ref{fig:log_cam_lz} and \ref{fig:ferrer_cam_lz}.

iii) To resolve finer orbital structure, the periodic orbits that define the boundaries of different families must be computed in the frozen potential. Chaotic orbits can be identified in the $\overline{L_z}$–$\sigma_{L_z}$ plane as regions of scattered points that deviate from the well-defined curves traced by quasi-periodic curves obtained from the iso-$E_\mathrm{J}$ sequence, as shown in the left panel of Figure~\ref{fig:ferrer_lzsigmalz_peri}.

%Before constructing the $\cam-\rms$ plane for the snapshot particles, one must first determine the locations of the periodic-orbit branches that delineate the regions associated with specific orbital families inside the corotation radius. These branches can be obtained using the gravitational potential computed directly from the snapshot. Once identified, orbits of snapshot particles can be classified by integrating them in this potential and mapping their trajectories onto the $\cam-\rms$ plane. This procedure is particularly effective for quasi-periodic orbits, which occupy substantial regions of the $\cam-\rms$ plane. 

%In contrast, chaotic orbits typically lie within a narrow band near $\cam \sim 0$, making them more readily identified in the $\overline{L_z}-\sigma_{L_z}$ plane. Because the distribution of simulated particles in the $\overline{L_z}-\sigma_{L_z}$ plane is dominated by orbits outside corotation, signatures of the bar are visible only in a zoomed-in region near the origin. As in the quasi-periodic case, the domain occupied by chaotic orbits can be isolated by computing iso-$E_\mathrm{J}$ sequences in the snapshot potential. Chaotic orbits then appear as scattered points that deviate from the well-defined curves traced by quasi-periodic families.

\section{Conclusion}
\label{section: conclusion}

In this paper, we propose a new method for 2D orbital classification by using a proxy of the second integral of motion, $\cam$, defined as the ratio of the time-averaged angular momentum to its temporal dispersion in the corotating frame. 

We begin by analytically examining the physical interpretation of $\cam$ within the Freeman bar model. Our analysis shows that $\overline{L_z}$ is dominated by the contribution from ${J_\phi}^\prime$, which describes the azimuthal motion, whereas $\sigma_{L_z}$ is largely governed by ${J_r}^\prime$, representing the radial motion. Consequently, $\cam$ depends on the ratio ${J_\phi}^\prime / {J_r}^\prime$ in the Freeman bar, offering valuable insight into its physical nature.

We construct a new parameter space defined by $\cam$ and the root-mean-square radius ($\rms$). We do not adopt the conventional integral-of-motion space defined by $\cam$ and $E_\mathrm{J}$, as it exhibits degeneracies arising from orbits outside the corotation radius.

We test our method in three rotating barred potentials: the Freeman bar, the logarithmic bar, and the Ferrers bar. We find that periodic orbits trace distinct branches in the $\cam-\rms$ plane, effectively partitioning it into regions occupied by their corresponding orbital families. Notably, some of these branches form closed areas that allow clear identification of specific families, such as the $x_2$ orbits. We further extend our analysis to test-particle orbits spanning a wide range of $E_\mathrm{J}$ and confirm that different orbital families exhibit a well-ordered distribution in this plane with no spatial overlap.

To conclude, we demonstrate that $\cam$ versus the root-mean-square radius plane provides an effective framework for orbital classification in rotating bar potentials. As CAM is fundamentally linked to the intrinsic properties of orbits, it is particularly well suited for application in $N$-body simulations and can be naturally extended to 3D cases. Although CAM may display ambiguous behavior for higher-order resonant or chaotic orbits, nonetheless it represents a valuable complement to conventional orbital classification methods, such as the SoS and frequency analysis method, owing to its simplicity and computational efficiency when applied to individual orbits.

\begin{acknowledgements}
We thank James Binney for useful comments. The research presented here:  is partially supported by the National Natural Science Foundation of China under grant Nos. 12533004, 12025302, 11773052; by China Manned Space Program with grant no. CMS-CSST-2025-A11; by the “111” Project of the Ministry of Education of China under grant No. B20019; and by Office of Science and Technology, Shanghai Municipal Government with grant Nos. 24DX1400100 and ZJ2023-ZD-001. This work has made use of the Gravity Supercomputer at the Department of Astronomy, Shanghai Jiao Tong University.
\end{acknowledgements}

\newpage
\appendix
 \renewcommand{\theequation}{\thesection.\arabic{equation}}

\section{$\cam-E_{\mathrm{J}}$ space for orbits inside the corotation radius}
\label{appendix:insideco}

\begin{figure*}[htbp!]
  \centering
  \includegraphics[width=0.49\textwidth]{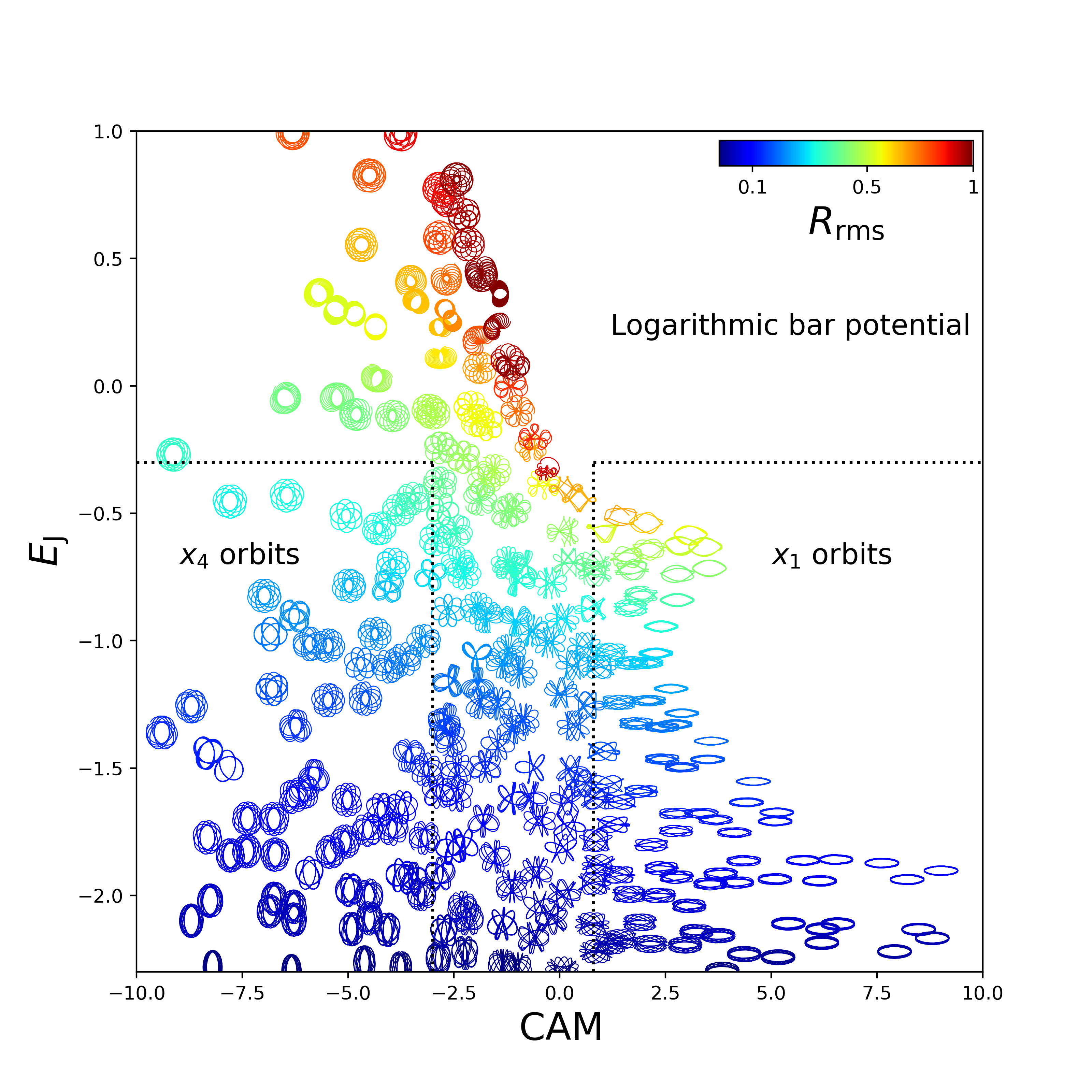}
  \includegraphics[width=0.49\textwidth]{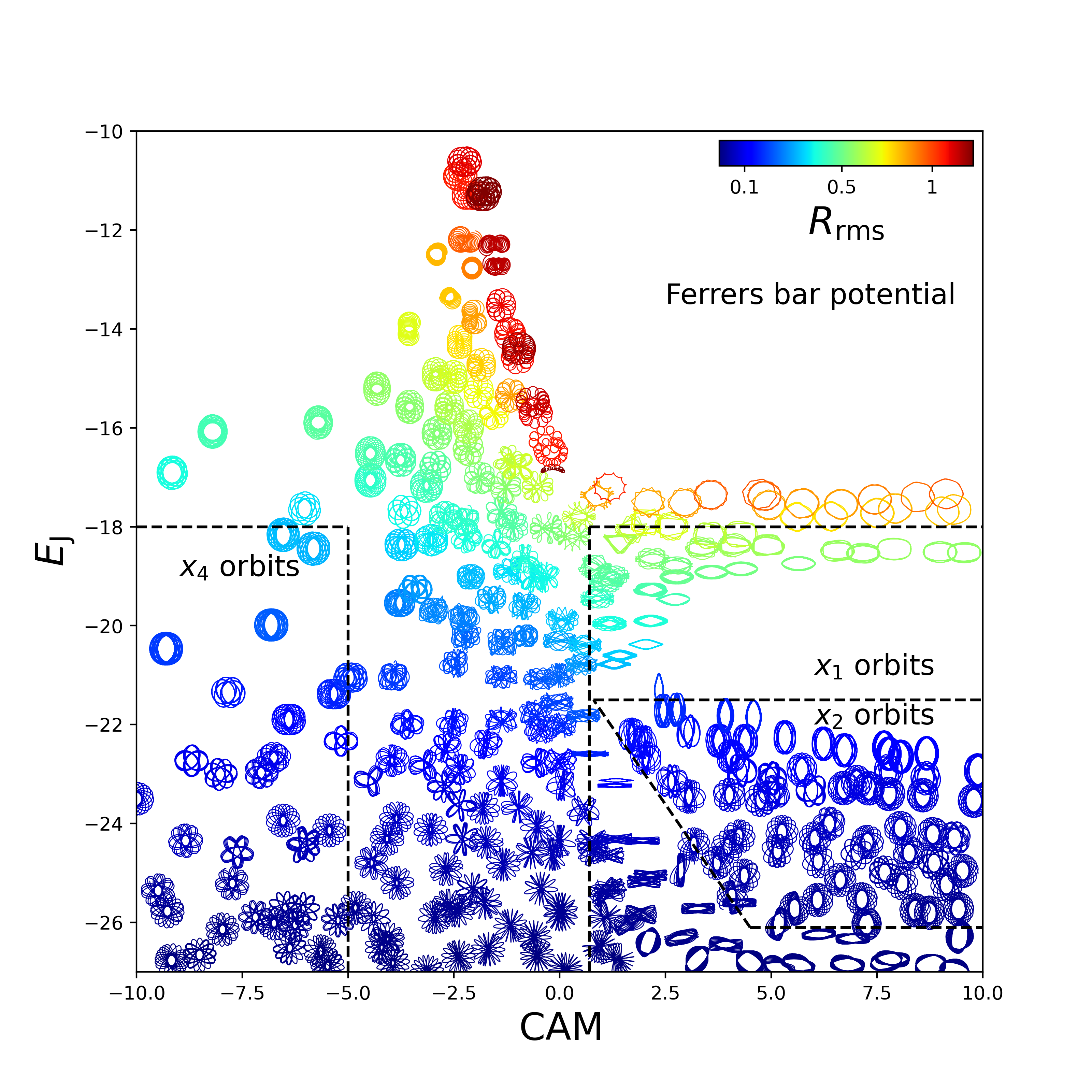}

  \caption{Distribution of orbits inside the corotation radius in the $\cam-E_\mathrm{J}$ plane for logarithmic bar potential (left) and Ferrers bar potential (right), color-coded by $\rms$. Scaled orbits are plotted at their corresponding locations in both panels. The approximate regions of the $x_1$, $x_2$ and $x_4$ orbital families are outlined by black dashed lines. }
  \label{fig:inside_co}
\end{figure*}

In this section, we examine how orbits inside the corotation radius of the logarithmic and Ferrers bar potentials are distributed in the $\cam-E_{\mathrm{J}}$ plane, as shown in Figure~\ref{fig:inside_co}. The $\cam-E_{\mathrm{J}}$ plane can be viewed as an alternative to the $\cam-\rms$ plane once all orbits outside corotation are removed. In both representations, bar-supporting orbits exhibit similar morphological distributions.

\newpage

\end{document}